\documentclass{lmcs} 
\pdfoutput=1

\usepackage[utf8]{inputenc}

\usepackage{lastpage}
\lmcsdoi{17}{3}{10}
\lmcsheading{}{\pageref{LastPage}}{}{}%
{Feb.~26,~2020}{Jul.~23,~2021}{}


\usepackage{mathpartir}
\usepackage{amsmath}
\usepackage{amssymb}
\usepackage{amsthm}
\usepackage{enumitem}
\usepackage{todonotes}
\usepackage{trsym}
\usepackage{color}
\usepackage{verbatim}
\usepackage{url}
\usepackage{wrapfig}
\usepackage{rotating}
\usepackage{booktabs}
\makeatletter
\def\url@leostyle{%
  \@ifundefined{selectfont}{\def\UrlFont{\sf}}{\def\UrlFont{\small\ttfamily}}}
\makeatother

\usepackage[curve]{xypic}

\usepackage{graphics}
\usepackage{graphicx}
\usepackage[all]{xy}
\usepackage{eucal}

\theoremstyle{plain} 

\newdimen\proofrulebreadth \proofrulebreadth=.05em
\newdimen\proofdotseparation \proofdotseparation=1.25ex
\newdimen\proofrulebaseline \proofrulebaseline=2ex
\newcount\proofdotnumber \proofdotnumber=3
\let\then\relax
\def\hfi{\hskip0pt plus.0001fil}
\mathchardef\squigto="3A3B
\newif\ifinsideprooftree\insideprooftreefalse
\newif\ifonleftofproofrule\onleftofproofrulefalse
\newif\ifproofdots\proofdotsfalse
\newif\ifdoubleproof\doubleprooffalse
\let\wereinproofbit\relax
\newdimen\shortenproofleft
\newdimen\shortenproofright
\newdimen\proofbelowshift
\newbox\proofabove
\newbox\proofbelow
\newbox\proofrulename
\def\shiftproofbelow{\let\next\relax\afterassignment\setshiftproofbelow\dimen0 }
\def\shiftproofbelowneg{\def\next{\multiply\dimen0 by-1 }%
\afterassignment\setshiftproofbelow\dimen0 }
\def\setshiftproofbelow{\next\proofbelowshift=\dimen0 }
\def\setproofrulebreadth{\proofrulebreadth}

\def\prooftree{
\ifnum  \lastpenalty=1
\then   \unpenalty
\else   \onleftofproofrulefalse
\fi
\ifonleftofproofrule
\else   \ifinsideprooftree
        \then   \hskip.5em plus1fil
        \fi
\fi
\bgroup
\setbox\proofbelow=\hbox{}\setbox\proofrulename=\hbox{}%
\let\justifies\proofover\let\leadsto\proofoverdots\let\Justifies\proofoverdbl
\let\using\proofusing\let\[\prooftree
\ifinsideprooftree\let\]\endprooftree\fi
\proofdotsfalse\doubleprooffalse
\let\thickness\setproofrulebreadth
\let\shiftright\shiftproofbelow \let\shift\shiftproofbelow
\let\shiftleft\shiftproofbelowneg
\let\ifwasinsideprooftree\ifinsideprooftree
\insideprooftreetrue
\setbox\proofabove=\hbox\bgroup$\displaystyle 
\let\wereinproofbit\prooftree
\shortenproofleft=0pt \shortenproofright=0pt \proofbelowshift=0pt
\onleftofproofruletrue\penalty1
}

\def\eproofbit{
\ifx    \wereinproofbit\prooftree
\then   \ifcase \lastpenalty
        \then   \shortenproofright=0pt  
        \or     \unpenalty\hfil         
        \or     \unpenalty\unskip       
        \else   \shortenproofright=0pt  
        \fi
\fi
\global\dimen0=\shortenproofleft
\global\dimen1=\shortenproofright
\global\dimen2=\proofrulebreadth
\global\dimen3=\proofbelowshift
\global\dimen4=\proofdotseparation
\global\count255=\proofdotnumber
$\egroup  
\shortenproofleft=\dimen0
\shortenproofright=\dimen1
\proofrulebreadth=\dimen2
\proofbelowshift=\dimen3
\proofdotseparation=\dimen4
\proofdotnumber=\count255
}

\def\proofover{
\eproofbit 
\setbox\proofbelow=\hbox\bgroup 
\let\wereinproofbit\proofover
$\displaystyle
}%
\def\proofoverdbl{
\eproofbit 
\doubleprooftrue
\setbox\proofbelow=\hbox\bgroup 
\let\wereinproofbit\proofoverdbl
$\displaystyle
}%
\def\proofoverdots{
\eproofbit 
\proofdotstrue
\setbox\proofbelow=\hbox\bgroup 
\let\wereinproofbit\proofoverdots
$\displaystyle
}%
\def\proofusing{
\eproofbit 
\setbox\proofrulename=\hbox\bgroup 
\let\wereinproofbit\proofusing
\kern0.3em$
}

\def\endprooftree{
\eproofbit 
  \dimen5 =0pt
\dimen0=\wd\proofabove \advance\dimen0-\shortenproofleft
\advance\dimen0-\shortenproofright
\dimen1=.5\dimen0 \advance\dimen1-.5\wd\proofbelow
\dimen4=\dimen1
\advance\dimen1\proofbelowshift \advance\dimen4-\proofbelowshift
\ifdim  \dimen1<0pt
\then   \advance\shortenproofleft\dimen1
        \advance\dimen0-\dimen1
        \dimen1=0pt
        \ifdim  \shortenproofleft<0pt
        \then   \setbox\proofabove=\hbox{%
                        \kern-\shortenproofleft\unhbox\proofabove}%
                \shortenproofleft=0pt
        \fi
\fi
\ifdim  \dimen4<0pt
\then   \advance\shortenproofright\dimen4
        \advance\dimen0-\dimen4
        \dimen4=0pt
\fi
\ifdim  \shortenproofright<\wd\proofrulename
\then   \shortenproofright=\wd\proofrulename
\fi
\dimen2=\shortenproofleft \advance\dimen2 by\dimen1
\dimen3=\shortenproofright\advance\dimen3 by\dimen4
\ifproofdots
\then
        \dimen6=\shortenproofleft \advance\dimen6 .5\dimen0
        \setbox1=\vbox to\proofdotseparation{\vss\hbox{$\cdot$}\vss}%
        \setbox0=\hbox{%
                \advance\dimen6-.5\wd1
                \kern\dimen6
                $\vcenter to\proofdotnumber\proofdotseparation
                        {\leaders\box1\vfill}$%
                \unhbox\proofrulename}%
\else   \dimen6=\fontdimen22\the\textfont2 
        \dimen7=\dimen6
        \advance\dimen6by.5\proofrulebreadth
        \advance\dimen7by-.5\proofrulebreadth
        \setbox0=\hbox{%
                \kern\shortenproofleft
                \ifdoubleproof
                \then   \hbox to\dimen0{%
                        $\mathsurround0pt\mathord=\mkern-6mu%
                        \cleaders\hbox{$\mkern-2mu=\mkern-2mu$}\hfill
                        \mkern-6mu\mathord=$}%
                \else   \vrule height\dimen6 depth-\dimen7 width\dimen0
                \fi
                \unhbox\proofrulename}%
        \ht0=\dimen6 \dp0=-\dimen7
\fi
\let\doll\relax
\ifwasinsideprooftree
\then   \let\VBOX\vbox
\else   \ifmmode\else$\let\doll=$\fi
        \let\VBOX\vcenter
\fi
\VBOX   {\baselineskip\proofrulebaseline \lineskip.2ex
        \expandafter\lineskiplimit\ifproofdots0ex\else-0.6ex\fi
        \hbox   spread\dimen5   {\hfi\unhbox\proofabove\hfi}%
        \hbox{\box0}%
        \hbox   {\kern\dimen2 \box\proofbelow}}\doll%
\global\dimen2=\dimen2
\global\dimen3=\dimen3
\egroup 
\ifonleftofproofrule
\then   \shortenproofleft=\dimen2
\fi
\shortenproofright=\dimen3
\onleftofproofrulefalse
\ifinsideprooftree
\then   \hskip.5em plus 1fil \penalty2
\fi
}

\DeclareMathOperator{\supp}{supp}

\newcommand{\id}{\mbox{\sl id}}
\newcommand{\after}{\mathrel{\circ}}
\newcommand{\calD}{\mbox{$\mathcal D$}}

\newcommand{\Pow}{\mathcal{P}}
\newcommand{\Powne}{\mathcal{P}_{\mathit{ne}}}

\newcommand{\idmap}{\textrm{id}}
\newcommand{\cat}[1]{\ensuremath{\mathbf{#1}}}
\newcommand{\Cat}[1]{\ensuremath{\mathbf{#1}}}
\newcommand{\Sets}{\Cat{Sets}}

\newcommand{\Kl}{\mathcal{K}{\kern-.2ex}\ell}

\newcommand{\set}[2]{\{#1\;|\;#2\}}
\newcommand{\setin}[3]{\{#1\in#2\;|\;#3\}}

\newcommand{\Alg}[1]{\mathbb{#1}}			
\DeclareMathOperator{\con}{conv}			    
\DeclareMathOperator{\cvx}{\con}			    
\DeclareMathOperator{\cgr}{cgr}			    
\DeclareMathOperator{\f}{f}			    
\DeclareMathOperator{\g}{g}			    
\DeclareMathOperator{\bis}{b}	
\DeclareMathOperator{\ide}{id}			    
\DeclareMathOperator{\rfl}{r}			    
\DeclareMathOperator{\sym}{s}			    
\DeclareMathOperator{\trans}{t}			    
\DeclareMathOperator{\e}{e}			    

\DeclareMathOperator{\CoAlg}{Coalg}
\DeclareMathOperator{\EM}{E{\kern-.2ex}M}

\newcommand{\PowC}{\mathcal{P}_{{\rm c}}}	
\newcommand{\Dis}{\mathcal{D}}				
\newcommand{\Definitive}[1]{(\PowC{#1}+1)^L}   
\newcommand{\Definit}{(\PowC+1)^L}   

\newcommand{\T}{\mathcal M}				
\newcommand{\High}{\mathcal{H}}				
\newcommand{\Lone}{\mathcal{L}_1}		
\newcommand{\Ltwo}{\mathcal{L}_2}		

\newcommand{\Ubar}{\overline{\mathcal{U}}}		
\newcommand{\GPC}{\overline{\mathcal{F}}} 
\newcommand{\Fbar}{\overline{\mathcal{F}}}

\newcommand{\psum}[3]{#1 #2 + \bar{#1} #3} 
\newcommand{\tr}[1]{ \stackrel{#1}{\to}}           
\newcommand{\Tr}[1]{ \stackrel{#1}{\mapsto}}           
\newcommand{\Btr}[1]{ \stackrel{#1}{\mapsto}}           
\newcommand{\BSTeq}{\sim_d}                             

\newcommand{\Act}{L}

\newcommand{\qedd}{\hfill\ensuremath{\diamond}}

\begin{document}

 
\title{Distribution Bisimilarity via\\ The Power of Convex Algebras}
 
\titlecomment{{\lsuper*}
This paper and the SI is dedicated to Jos Baeten---a pioneer of formal methods and concurrency theory, and a strong supporter of algebraic methods in computer science. We highly respect Jos' scientific work and his personal integrity, and thank him for his contributions to the whole area and to our work in particular. Almost exactly 20 years ago, I (Ana Sokolova) have joined Jos' Formal Methods Group at the TU Eindhoven as a PhD student. I am utterly grateful for Jos' guidance, sense of justice, work ethics, and for teaching me to appreciate beauty, freedom, and independence in Science. We share Jos' love for algebra and fascination with probabilistic systems, as it is evident from this paper.
The paper is an extended version of the paper~\cite{BSS17} published at CONCUR 2017. The current paper contains all proofs and many additional remarks, clarifications, and examples. Bonchi's work was supported by the Ministero dell'Universit\`a e della Ricerca of Italy under Grant Nr. 201784YSZ5, PRIN2017 - ASPRA (Analysis of Program Analyses). Silva's work was partially supported by the ERC Consolidator Grant AutoProbe (grant code 101002697) and a Royal Society Wolfson Fellowship.}

\author[F.~Bonchi]{Filippo Bonchi}	
\address{University of Pisa, Italy, and CNRS, ENS-Lyon, France}	
\email{filippo.bonchi@ens-lyon.fr}  

\author[A.~Silva]{Alexandra Silva}	
\address{University College London, UK}	
\email{alexandra.silva@ucl.ac.uk}  

\author[A.~Sokolova]{Ana Sokolova}	
\address{University of Salzburg, Austria}	
\email{ana.sokolova@cs.uni-salzburg.at}




%
%
%



\begin{abstract}
Probabilistic automata (PA), also known as probabilistic nondeterministic labelled transition systems, combine probability and nondeterminism. They can be given different semantics, like strong bisimilarity, convex bisimilarity, or  (more recently) distribution bisimilarity. The latter is based on the view of PA as transformers of probability distributions, also called belief states, and promotes distributions to first-class citizens.  

We give a coalgebraic account of distribution bisimilarity, and explain the genesis of the belief-state transformer from a PA. 
To do so, we make explicit the convex algebraic structure present in PA and identify belief-state transformers as transition systems with state space that carries a convex algebra. As a consequence of our abstract approach, we can give a sound proof technique which we call bisimulation up-to convex hull.
\end{abstract}

\maketitle

\section{Introduction}\label{sec:intro}
Probabilistic automata (PA), closely related to Markov decision processes (MDPs), have been used along the years in various areas of verification~\cite{DBLP:conf/tacas/LegayMOW08,DBLP:conf/cav/KwiatkowskaNP11,DBLP:journals/tcs/KwiatkowskaNSS02,Baier2008}, machine learning~\cite{DBLP:journals/corr/abs-1206-3255,DBLP:journals/ml/MaoCJNLN16}, and semantics~\cite{DBLP:journals/corr/abs-1305-5055,DBLP:conf/ifipTCS/SherK12}. Recent interest in research around semantics of probabilistic programming languages has led to new insights in connections between category theory, probability theory, and automata~\cite{DBLP:conf/lics/StatonYWHK16,DBLP:conf/concur/DahlqvistDG16,DBLP:journals/corr/HeunenKSY17,DBLP:conf/esop/Staton17,DBLP:conf/birthday/Mislove17}. 

 PA have been given various semantics, starting from strong bisimilarity~\cite{LS91:ic}, probabilistic (convex) bisimilarity~\cite{SL94,Seg95:thesis}, to bisimilarity on distributions~\cite{DoyenHR08,DGHM08,CPP09,EisentrautHZ10,CR11,Hennessy12,FengZ14,HermannsKK14}. In this last view, probabilistic automata are understood as transformers of belief states, labeled transition systems (LTSs) having as states probability distributions, see 
 e.g.~\cite{DGHM08,DengGHM09,KorthikantiVAK10,AgrawalAGT12,DengH13,FengZ14,DoyenMS14}. Checking such equivalence raises a lot of challenges since belief states are uncountable. Nevertheless, it is  decidable~\cite{HermannsKK14,EisentrautHKT013} with help of convexity. Despite these developments, what remains open is the understanding of the genesis of belief-state transformers and canonicity of distribution bisimilarity, as well as the role of  convex algebras. This paper provides the answers and explains the belief-state transformers and distribution bisimilarity with the help of coalgebraic techniques.

The theory of coalgebras~\cite{J2016,Rut00:tcs,JR96:eatcs} provides a toolbox for modelling and analysing different types of state machines. In a nutshell, a coalgebra is an arrow $c\colon S\to FS$ for some functor $F\colon \cat{C} \to \cat{C}$ on a category $\cat{C}$. Intuitively, $S$ represents the space of states of the machine, $c$ its transition structure and the functor $F$ its type. Most importantly, every functor gives rise to a canonical notion of behavioural equivalence ($\approx$), a coinductive proof technique and, for finite states machines, a procedure to check $\approx$. 

By tuning the parameters $\cat{C}$ and $F$, one can retrieve many existing types of machines and their associated equivalences. For instance, by taking $\cat{C}= \Sets$, the category of sets and functions, and $FS = (\Pow \Dis S)^\Act$, the set of functions from $\Act$ to subsets ($\Pow$) of probability distributions ($\Dis$) over $S$, coalgebras $c\colon S\to FS$ are in one-to-one correspondence with PA with labels in $L$. Moreover, the associated notion of behavioural equivalence turns out to be the classical strong probabilistic bisimilarity of~\cite{LS91:ic} (see~\cite{BSV04:tcs,Sokolova11} for more details). Recent work~\cite{Mio14} shows that, by taking a slightly different functor, forcing the subsets to be convex, one obtains probabilistic (convex) bisimilarity as in~\cite{SL94,Seg95:thesis}.

In this paper, we take a coalgebraic outlook at the semantics of probabilistic automata as belief-state transformers: we wish to translate a PA $c\colon S \to  (\Pow \Dis S)^\Act$ into a belief-state transformer $c^\sharp \colon \Dis S \to (\Pow \Dis S)^\Act$. Note that the latter is a coalgebra for the functor $FX=(\Pow X)^\Act$, i.e., a labeled transition system, since the state space is the set of probability distributions $\Dis S$. This is reminiscent of the standard determinisation for non-deterministic automata (NDA) seen as coalgebras $c\colon S\to 2\times (\Pow S)^L$. The result of the determinisation is a deterministic automaton $c^\sharp\colon \Pow S \to 2\times (\Pow S)^L$ (with state space $\Pow S$), which is a coalgebra for the functor $FX = 2\times X^L$. In the case of PA, one lifts the states space to $\Dis S$, in the one of NDA to $\Pow S$. From an abstract perspective, both $\Dis$ and $\Pow$ are monads, hereafter denoted by $\T$, and both PA and NDA can be regarded as coalgebras of type $c\colon S \to F\T S$. 

In \cite{SBBR10}, a generalised determinisation transforming coalgebras $c\colon S \to F\T S$ into coalgebras $c^\sharp \colon \T S \to F\T S$ was presented. This construction requires the existence of a \emph{lifting} $\overline{F}$ of $F$ to the category of algebras for the monad $\T$. In the case of NDA, the functor $FX = 2\times X^L$ can be easily lifted to the category of join-semilattices (algebras for $\Pow$) and the coalgebra $c^\sharp\colon \Pow S \to 2\times (\Pow S)^L$ resulting from this construction turns out to be exactly the standard determinised automaton. Unfortunately, this is not the case with probabilistic automata: because of the lack of a suitable distributive law of $\Dis$ over $\Pow$ \cite{VW06}, it is impossible to suitably lift $FX=(\Pow X)^\Act$ to the category of \emph{convex algebras} (algebras for the monad $\Dis$).

The way out of the impasse consists in defining a powerset-like functor on the category of convex algebras. This is not a functor lifting to the category of algebras but it enjoys enough properties that allow to transform every PA into a labelled transition system on convex algebras. In turn, these can be transformed---without changing the underlying behavioural equivalence---into standard LTSs on $\Sets$ by simply forgetting the algebraic structure. 
We show that the result of the whole procedure is exactly the expected belief-state transformer and that the induced notion of behavioural equivalence coincides with a canonical distribution bisimilarity present in the literature~\cite{DGHM08,Hennessy12,FengZ14,HermannsKK14}.

Recognising the analogy between  this new construction and the above on NDA pays off in terms of proof techniques. In \cite{bonchi2013checking}, Bonchi and Pous introduced an efficient algorithm to check language equivalence of NDA based on coinduction up-to~\cite{PS11}: in a determinised automaton $c^\sharp\colon \Pow S \to 2\times (\Pow S)^L$, language equivalence can be proved by means of bisimulations  up-to the structure of join semilattice carried by the state space $\Pow S$. Algorithmically, this results in an impressive pruning of the search space.

Similarly, in a belief-state transformer $c^\sharp \colon \Dis S \to (\Pow \Dis S)^\Act$, one can coinductively reason up-to the convex algebraic structure carried by $\Dis S$. The resulting proof technique, which we call in this paper \emph{bisimulation up-to convex hull}, allows finite relations to witness the equivalence of infinitely many states.
More precisely, by exploiting a recent result in convex algebra by Sokolova and Woracek~\cite{SW15}, we are able to show that distribution bisimilarity of any two belief states can always be proven by means of a \emph{finite} bisimulation up-to.

The paper starts with background on PA (Section~\ref{sec:PA}), convex algebras (Section~\ref{sec:CA}), and coalgebra (Section~\ref{sec:coalgebra}). We provide the PA functor on convex algebras in Section~\ref{sec:IAPAonCA}. We give the transformation from PA to belief-state transformers in Section~\ref{sec:Landscape} and prove the coincidence of the abstract and concrete transformers and (distribution bisimilarity) semantics. We present bisimulation up-to convex hull and prove soundness in Section~\ref{sec:bis-upto-cc}. 


\section{Probabilistic Automata}\label{sec:PA}

\newcommand*{\Cdot}{\raisebox{-0.25ex}{\scalebox{1.4}{$\cdot$}}}
\begin{figure}\scalebox{1}{
 $ \begin{array}{ccc|ccc}
             \xymatrix @C=1.8em @R=2.3em { %
      && & x_3  \\ 
      &&\cdot \ar@{.>}[ru]|(0.4){1} & \\
          & x_0\ar[r]^{a} \ar@/^/[ur]^{a} &\cdot \ar@{.>}[r]|(0.4){1} & %
          x_1 \ar[r]^{a} \ar@/_/[d]_{a} & \cdot \ar@{.>}[luu]|(0.2){\frac{1}{2}} \ar@{.>}[ldd]|(0.2){\frac{1}{2}} 
      \\    
      &  &  & \cdot \ar@{.>}[d]|(0.35){\frac{1}{2}}  \ar@/_/@{.>}[u]|(0.35){\frac{1}{2}}
      \\
      &&\cdot \ar@/^/@{.>}[r]|(0.4){1}& x_2 \ar@/^/[l]^{a}  }
      & \hspace*{-5mm}
             \xymatrix @C=1.8em @R=2em { %
      && & y_3  \\ 
      &&\cdot \ar@{.>}[ru]|(0.4){1} & \cdot \ar@{.>}[u]|(0.4){1} \\
          & y_0\ar@/^/[ur]_{a} \ar@/_/[dr]^{a} & & %
          y_1 \ar[u]^a \ar@/_/[d]^{a} & 
      \\    
      &  & \cdot \ar@{.>}[ru]|(0.3){\frac{1}{2}} \ar@{.>}[rd]|(0.3){\frac{1}{2}} & \cdot \ar@{.>}[d]|(0.35){\frac{1}{2}}  \ar@/_/@{.>}[u]|(0.35){\frac{1}{2}}
      \\
      &&\cdot \ar@/^/@{.>}[r]|(0.4){1}& y_2 \ar@/^/[l]^{a}  }
      &&&
      \xymatrix @C=1.9em @R=1.5em { %
      y_0 \ar@{|->}[d]_a="x" \ar@{|->}[dr]^a\ar@{}[dr]_(0.4){}="y"   \\ 
     {\frac{1}{2}y_1 + \frac{1}{2}y_2} \ar@{|->}[d]_a="x1" \ar@{|->}[dr]^a\ar@{}[dr]_(0.4){}="y1"& y_3 \\
      {\frac{1}{4}y_1 + \frac{3}{4}y_2}  \ar@{|->}[d]_a="x2" \ar@{|->}[dr]^a \ar@{}[dr]_(0.4){}="y2"&  {\frac{1}{2}y_3 + \frac{1}{2}y_2} \\
      {\frac{1}{8}y_1 + \frac{7}{8}y_2} \ar@{|->}[d]_a="x3" \ar@{|->}[dr]^a \ar@{}[dr]_(0.4){}="y3"&  {\frac{1}{4}y_3 + \frac{3}{4}y_2} \\
\dots & {\frac{1}{8}y_3 + \frac{7}{8}y_2}
\ar@{{}*\composite{{}*{\Cdot}}}@/_.8pc/"x";"y"
\ar@{{}*\composite{{}*{\Cdot}}}@/_.8pc/"x1";"y1"
\ar@{{}*\composite{{}*{\Cdot}}}@/_.8pc/"x2";"y2"
\ar@{{}*\composite{{}*{\Cdot}}}@/_.8pc/"x3";"y3"
}
    \end{array}$}
\caption{On the left: a PA with set of actions $\Act = \{a\}$ and set of states $S = \{x_0,\,x_1,\,x_2,\,x_3,\,y_0,\,y_1,\,y_2,\,y_3\}$.  We  depict each transition $s\tr{a} \zeta$ in two stages: a straight action-labeled arrow from $s$ to $\cdot$ and then several dotted arrows from $\cdot$ to  states in $S$ specifing the distribution $\zeta$. On the right: part of the corresponding belief-state transformer. The dots between two arrows $\zeta\Btr{a}\xi_1$ and $\zeta\Btr{a}\xi_2$ denote that $\zeta$ can perform infinitely many transitions to states obtained as convex combinations of $\xi_1$ and $\xi_2$. For instance $y_0 \Btr{a}\frac{1}{4}y_1 + \frac{1}{4}y_2 + \frac{1}{2}y_3$.}
  \label{fig:weighted}
\end{figure}
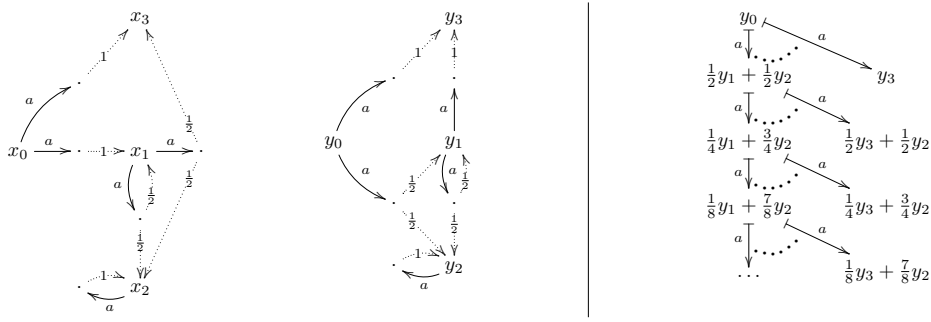
Probabilistic automata are models of systems that involve both probability and nondeterminism. We start with their definition  by Segala and Lynch~\cite{SL94}.

\begin{defi}\label{def:PA-concrete} 	A probabilistic automaton (PA) is a triple $M = (S, \Act, \to)$ where $S$ is a set of states, $\Act$ is a set of actions or action labels, and $\to \,\,\subseteq \,\,S \times \Act \times \Dis (S)$ is the transition relation. As usual, $s \stackrel{a}{\to} \zeta$  stands for $(s,a,\zeta) \in \,\,\to$. \qedd
\end{defi}
An example is shown on the left of Figure ~\ref{fig:weighted}. 	
Probabilistic automata can be given different semantics, e.g., (strong probabilistic) bisimilarity~\cite{LS91:ic}, convex (probabilistic) bisimilarity~\cite{SL94}, and as transformers of belief states~~\cite{CPP09,FengZ14,DengH13,DengGHM09,DGHM08,HermannsKK14} whose definitions we present next. For the rest of the section, we fix a PA $M = (S, \Act, \to)$.

\begin{defi}[Strong Probabilistic Bisimilarity]\label{def:bisim}
  	A relation $R \subseteq S \times S$ is a (strong probabilistic) \emph{bisimulation} if 
  	$(s,t) \in R$ implies, for all actions $a \in \Act$ and all $\xi, \zeta \in \Dis(S)$, that
\begin{eqnarray*}
s \stackrel{a}{\to} \xi  \Rightarrow  \exists \xi' \in \Dis(S). \,\, t \stackrel{a}{\to} \xi' \wedge \xi \equiv_R \xi', \quad \text{and} \quad 
t \stackrel{a}{\to} \zeta \Rightarrow \exists \zeta' \in \Dis(S). \,\, s \stackrel{a}{\to} \zeta' \wedge \zeta' \equiv_R \zeta.
\end{eqnarray*}

Here, $\equiv_R \,\,\subseteq\,\, \Dis(S)\times \Dis(S)$ is the lifting of $R$ to distributions, defined by $\xi \equiv_R \xi'$ if and only if there
exists a distribution $\nu \in \calD(S\times S)$ such that
\begin{itemize}[itemsep=1mm]
\item[1.] $\sum_{t \in S} \nu(s,t) = \xi(s)$ for any $s \in S$,
\item[2.] $\sum_{s \in S} \nu(s,t)
 = \xi'(t)$ for any $t \in T$, and  
 \item[3.] $\nu(s,t) \ne 0$ implies
$(s,t) \in R$.
\end{itemize} 
Two states $s$ and $t$ are (strongly probabilistically) \emph{bisimilar}, notation $s \sim t$, if there exists a (strong probabilistic) bisimulation $R$ with $(s,t) \in R$. \qedd
\end{defi}

\begin{defi}[Convex Bisimilarity]\label{def:c-bisim}
  	A relation $R \subseteq S \times S$ is a \emph{convex} (probabilistic) \emph{bisimulation} if 
  	$(s,t) \in R$ implies, for all actions $a \in \Act$ and all $\xi, \zeta \in \Dis(S)$, that
\begin{eqnarray*}
s \stackrel{a}{\to} \xi  \Rightarrow  \exists \xi' \in \Dis(S). \,\, t \stackrel{a}{\to}_c \xi' \wedge \xi \equiv_R \xi',\quad \text{and} \quad 
t \stackrel{a}{\to} \zeta  \Rightarrow  \exists \zeta' \in \Dis(S). \,\, s \stackrel{a}{\to}_c \zeta' \wedge \zeta' \equiv_R \zeta.
\end{eqnarray*}

Here $\to_c$ denotes the convex transition relation, defined as follows: $s  \stackrel{a}{\to}_c \xi$  if and only if  
$\xi = \sum_{i=1}^n p_i\xi_i$ for some $\xi_i \in \Dis(S)$ and $p_i \in [0,1]$ satisfying $\sum_{i=1}^n p_i = 1$ and $s \stackrel{a}{\to} \xi_i$ for $i = 1, \dots, n$.
Two states $s$ and $t$ are \emph{convex bisimilar}, notation $s \sim_c t$, if there exists a convex bisimulation $R$ with $(s,t) \in R$. \qedd
\end{defi}

Convex bisimilarity is (strong probabilistic) bisimilarity on the ``convex closure" of the given PA. More precisely, consider the PA $M_c \,= \, (S, L, \to_c)$ in which $s \stackrel{a}{\to}_c \xi$ whenever $s \in S$ and $\xi$ is in the convex hull (see Section~\ref{sec:CA} for a definition) of the set $\{\zeta \in \Dis(S) \mid s \stackrel{a}{\to} \zeta\}$. Then convex bisimilarity of $M$ is bisimilarity of $M_c$. 
Hence, if bisimilarity is the behavioural equivalence of interest, we see that convex semantics arises from a different perspective on the representation of a PA: instead of seeing the given transitions as independent, we look at them as generators of infinitely many transitions in the convex closure. For readers familiar with the schedulers, convex bisimilarity means that schedulers in PA can randomise.  

There is yet another way to understand PA, as belief-state transformers, present but sometimes implicit in~\cite{CPP09,Hennessy12,FengZ14,DengH13,DengGHM09,DGHM08,HermannsKK14,CR11} to name a few, with behavioural equivalences on distributions. We were particularly inspired by the original work of Deng et al.~\cite{DengH13,DengGHM09,DGHM08} as well as~\cite{HermannsKK14}.
Given a PA $M = (S, \Act, \to)$, consider the labeled transition system $M_{bs} = (\Dis S, \Act, \Tr{})$ with states distributions over the original states of $M$, and transitions $ \Tr{} \,\,\subseteq\,\, \Dis S \times \Act \times \Dis S$ defined by
$$ \xi \Tr{a} \zeta \quad \text{iff} \quad \xi = \sum p_i s_i, \,\, s_i \tr{a}_c \xi_i, \,\, \zeta = \sum p_i \xi_i.$$
We call $M_{bs}$ the belief-state transformer of $M$. Figure~\ref{fig:weighted}, right, displays a part of the belief-state transformer induced by the PA of Figure~\ref{fig:weighted}, left. According to this definition, a distribution makes an action step only if all its support states can make the step. 

This, and hence the corresponding notion of bisimulation, can vary. For example, in~\cite{HermannsKK14} a distribution makes a transition $\Btr{a}$ if some of its support states can perform an $\Btr{a}$ step
. 
There are several proposed notions of equivalences on distributions~\cite{Hennessy12,DoyenHR08,DoyenMS14,FengZ14,DengH13,CPP09,HermannsKK14}  that mainly differ in the treatment of termination
. See~\cite{HermannsKK14} for a detailed comparison. 

In this paper we focus on \emph{full} probability distributions for reasons of clarity and canonical presentation. This has some effect on the obtained semantics (and can lead to undesired corner cases). However, all our results can be recast in the setting of subprobability distributions too, yielding a more natural semantics on the expense of more complicated presentation. 

\begin{defi}[Distribution Bisimilarity]\label{def:distr-bisim}    An equivalence $R \subseteq \Dis S \times \Dis S$ is a distribution bisimulation of $M$ if and only if it is a bisimulation of the belief-state transformer $M_{bs}$. 	
    
    Two distributions $\xi$ and $\zeta$ are \emph{distribution bisimilar}, notation $\xi \sim_d \zeta$, if there exists a bisimulation $R$ with $(\xi,\zeta) \in R$. 
    Two states $s$ and $t$ are \emph{distribution bisimilar}, notation $s \sim_d t$, if $\delta_s \sim_d \delta_t$, where $\delta_x$ denotes the Dirac distribution with $\delta_x(x) = 1$. \qedd
\end{defi}

While the foundations of strong probabilistic bisimilarity are well-studied~\cite{Sokolova11,BSV04:tcs,VR99:tcs} and convex probabilistic bisimilarity was also recently captured coalgebraically~\cite{Mio14}, the foundations of the semantics of PA as transformers of belief states is  not yet explained. One of the goals of the present paper is to show that also that semantics (naturally on distributions~\cite{HermannsKK14}) is an instance of generic behavioural equivalence. Note that a (somewhat concrete) proof is given for the bisimilarity of~\cite{HermannsKK14}---the authors have proven that their bisimilarity is coalgebraic bisimilarity of a certain coalgebra corresponding to the belief-state transformer. What is missing there, and in all related work, is an explanation of the relationship of the belief-state transformer to the original PA. Clarifying the foundations of the belief-state transformer and the distribution bisimilarity is our initial motivation. 

A useful side effect of such clarification is the proof technique that we name up-to convex hull: the state $y_0$ of the belief-state transformer on the right of Figure \ref{fig:weighted} can reach infinitely many states (e.g., the distributions $\frac{1}{2^n} y_1 + (1 -\frac{1}{2^n}) y_2$ for all natural numbers $n$); therefore checking whether $y_0$ is distribution bisimilar to another state, e.g., $x_0$, would require a distribution bisimulation which relates infinitely many states. However, as we shall see in Example \ref{exuptocvx}, a \emph{finite} relation which is a bisimulation up-to convex hull will be enough to prove that $y_0$ is distribution bisimilar to $x_0$. Intuitively, bisimulations up-to convex hull allow to exploit the algebraic structure underlying probability distributions. We illustrate this structure in the next section.

\section{Convex Algebras} \label{sec:CA}
In order to model belief-state transformers as coalgebras, we first need to uncover the algebraic theory underlying probability distributions. By algebraic theory here we mean a set of operations (signature) and a set of equations that characterises, in a sense that we will make precise in Proposition~\ref{prop:EMD-CA}, probability distributions. It is well known that such theory is the one of convex algebras, that we quickly recall in this section.

By $\mathcal{C}$ we denote the signature of convex algebras
$$\mathcal{C} = \{ (p_i)_{i=0}^n \mid n \in \mathbb{N}, p_i \in [0,1], \sum_{i=0}^n p_i = 1 \}.$$
The operation symbol $(p_i)_{i=0}^n$ has arity $(n+1)$ and it will be interpreted by a convex combination with coefficients $p_i$ for $i=0, \dots, n$. 
For $p \in [0,1]$ we write $\bar{p} = 1-p$. 

\begin{defi}\label{def:CA}
	A \emph{convex algebra} $\Alg{X}$ is an algebra with signature $\mathcal{C}$, i.e., a set $X$ together with an 
	operation $\sum_{i=0}^n p_i (-)_i$ for each operational symbol $(p_i)_{i=0}^n \in \mathcal{C}$, 
	such that the following two axioms hold:
	\begin{itemize}[itemsep=1mm]
		\item Projection: $\sum_{i=0}^n p_ix_i = x_j$ if $p_j = 1$.
		\item Barycenter: $\sum_{i=0}^n p_i \left(\sum_{j=0}^m q_{i,j} x_j\right) = 
			\sum_{j=0}^m \left( \sum_{i=0}^n p_i q_{i,j}\right) x_j$. 
	\end{itemize}
A convex algebra homomorphism $h$ from $\Alg{X}$ to $\Alg{Y}$ is a \emph{convex} (synonymously, \emph{affine}) map, 
i.e., $h\colon X \to Y$ with the property $h\left(\sum_{i=0}^n p_i x_i \right) = \sum_{i=0}^n p_i h(x_i)$. \qedd
\end{defi}

\begin{rem}
Let $\Alg{X}$ be a convex algebra. Then (for $p_n \neq 1$)
\begin{equation}\label{eq:bin-suff}
	\sum_{i=0}^n p_i x_i = \overline{p_n} \left( \sum_{j=0}^{n-1} \frac{p_j}{\overline{p_n}} x_j\right) + p_n x_n
\end{equation}	
Hence, an $(n+1)$-ary convex combination can be written as a binary convex combination using an $n$-ary convex combination. As a consequence, if $X$ is a set that carries two convex algebras $\Alg{X}_1$ and $\Alg{X}_2$ with operations $\sum_{i=0}^n p_i (-)_i$ and $\bigoplus_{i=0}^n p_i (-)_i$, respectively (and binary versions $+$ and $\oplus$, respectively) such that $px + \bar p y = px \oplus \bar p y$ for all $p, x, y$, then $\Alg X_1 = \Alg X_2$. 
\end{rem}

One can also see~\eqref{eq:bin-suff} as a definition, see e.g.~\cite[Definition~1]{stone:1949}.  
We make the connection explicit with the next proposition, cf.\ \cite[Lemma~1-Lemma~4]{stone:1949}\footnote{Stone's cancellation Postulate~V is not used in his Lemma~1-Lemma~4.}. 

\begin{prop}\label{prop:bin-suff}
	Let $X$ be a set with binary operations $px + \bar py$ for $x, y \in X$ and $p \in (0,1)$. 
	For $x, y, z \in X$ and $p,q \in (0,1)$, assume 
	\begin{itemize}[itemsep=1mm]
		\item Idempotence: $px + \bar p x = x$,
		\item Parametric commutativity: $px + \bar py = \bar p y + p x$,
		\item Parametric associativity: $p(qx + \bar qy) + \bar p z = 
			pqx + \overline{pq}\left( \frac{p\bar q}{\overline{pq}}y + \frac{\bar p}{\overline{pq}}z\right)$,  
	\end{itemize}
	 and define $n$-ary convex operations by the projection axiom and the formula~\eqref{eq:bin-suff}. Then $X$ becomes a convex algebra.\qed
\end{prop}

Hence, it suffices to consider binary convex combinations only, whenever more convenient. 

\begin{defi}\label{def:convex-set}	Let $\Alg X$ be a convex algebra, with carrier $X$ and $C \subseteq X$. $C$ is \emph{convex} if it is the carrier of a subalgebra of $\Alg X$, i.e., if 
			$px + \bar p y \in C$ for $x, y \in C$ and $p \in (0,1)$. The \emph{convex hull} of a set $S \subseteq X$, denoted $\con(S)$, is the smallest convex set that contains $S$.  \qedd
\end{defi}
	
Clearly, a set $C \subseteq X$ for $X$ being the carrier of a convex algebra $\Alg X$ is convex if and only if $C = \con(C)$.  Convexity plays an important role in the semantics of probabilistic automata, for example in the definition of convex bisimulation, Definition~\ref{def:c-bisim}.  	

\begin{exa}\label{ex:convexalgebra}
Let $X$ be the set $\{x,y,z\}$ and $\Dis(X)$ the set of probability distributions over $X$, that is $\set{\varphi\colon X \to [0,1]}{\sum_{u \in X} \varphi(u) = 1}$. The latter set carries the structure of a convex algebra: for each $p\in (0,1)$, $\varphi_1, \varphi_2 \in \Dis(X)$, one can define $p\varphi_1 + \bar p \varphi_2$ as the function mapping each $u\in X$ into  $p\varphi_1(u) + \bar p \varphi_2(u)$. It is immediate to check that $p\varphi_1 + \bar p \varphi_2$ is indeed a distribution, since $\bar p = 1-p$, and that the thee equations of Proposition~\ref{prop:bin-suff} hold.

In order to illustrate the concept of convexity, we consider now some subsets of $\Dis(X)$. Take $C_1=\{\delta_x, \delta_y\}$ where $\delta_x$ and $\delta_y$ are the Dirac distributions for $x$ and $y$, respectively. This set is not convex since $\frac{1}{2} \delta_x + \frac{1}{2} \delta_y$,  the distribution mapping both $x$ and $y$ to $\frac{1}{2}$, does not belong to $C_1$. The smallest convex set that contains $C_1$ is its convex hull $\con(C_1)$: this is the set of all probability distributions over $\{x,y\}$.

For a similar reason, the set $C_2=\{\delta_x, \frac{1}{2} \delta_x + \frac{1}{2} \delta_y\}$ is not convex. The reader can easily check that its convex hull is $\con(C_2) = \set{\varphi\in \Dis(X)}{\varphi(x)\geq \frac{1}{2},\; \varphi(y)=1-\varphi(x)}$. 

In contrast, the set $C_3=\{\delta_x\}$ is convex. Indeed, for all $p\in (0,1)$, $p\delta_x + \bar p \delta_x$ is, by idempotence, equal to $\delta_x$. Finally, note that by definition the empty set is also convex.
\end{exa}


\section{Coalgebras and Generalised Determinisation}\label{sec:coalgebra}
%
In this section we give a gentle introduction to (co)algebra that enables us to highlight the generic principles behind the semantics of probabilistic automata. 
The interested reader is referred to~\cite{J2016,Rut00:tcs,JR96:eatcs} for more details.
We start by recalling even the basic notions of category, functor and natural transformation, so that all of the results in the paper are accessible also to non-experts.

A category $\cat C$ is a collection of objects and a collection of arrows (or morphisms) from one object to another. For every object $X \in \cat C$, there is an identity arrow $\id_X\colon X \to X$. For any three objects $X, Y, Z \in \cat C$, given two arrows $f\colon X \to Y$ and $g\colon Y\to Z$, there exists an arrow $g \after f\colon X \to Z$. Arrow composition is associative and $\id_X$ is neutral w.r.t. composition. 
The standard example is $\Sets$, the category of sets and functions. We will sometimes use the epi-mono factorisation of arrows in $\Sets$---every function is the composition of an epi followed by a mono---where an epi is a  surjective function and a mono is an injective function in $\Sets$. 

A functor $F$ from a category $\cat C$ to a category $\cat D$, notation $F\colon \cat C \to \cat D$, assigns to every object $X\in \cat C$, an object $FX \in \cat D$, and to every arrow $f\colon X \to Y$ in $\cat C$ an arrow $Ff\colon FX \to FY$ in $\cat D$ such that identity arrows and composition are preserved. 

\begin{exa} Examples of functors on $\Sets$ which are of interest to us are:
\begin{enumerate}
\item The constant exponent functor $(-)^L$ for a set $L$, mapping a set $X$ to the set $X^L$ of all functions from $L$ to $X$, and a function $f\colon X\to Y$ to $f^L\colon X^L \to Y^L$ with $f^L(g) = f \after g$.
\item The powerset functor $\Pow$ mapping a set $X$ to its powerset $\Pow X = \{S \mid S \subseteq X\}$ and on functions $f\colon X\to Y$ given by direct image: $\Pow f\colon \Pow X\to \Pow Y$, $\Pow(f)(S) = \{f(s) \mid s\in S\}$.
\item The finitely supported probability distribution functor $\Dis$ is defined, for a set $X$ and a function $f\colon X \to Y$,  as
\begin{eqnarray*}
\!\!\!\!\!\!\!\!
\Dis X
\,\,\, &=& \,\,\,
\set{\varphi\colon X \to [0,1]}{\sum_{x \in X} \varphi(x) = 1,\, \supp(\varphi) \text{~is~finite}}\\
\qquad \Dis f(\varphi)(y)
\,\,\, &=& \,\,\,
\sum\limits_{x\in f^{-1}(y)} \varphi(x).
\end{eqnarray*} 
\noindent The support set of a distribution $\varphi \in \Dis X$ is
defined as $$\supp(\varphi) = \setin{x}{X}{\varphi(x) \neq 0}.$$ 
\item The functor $C$~\cite{Mio14,Jacobs08,Varacca03}  maps a set $X$ to the set of all nonempty convex subsets of distributions over $X$, and a function $f\colon X \to Y$ to the function $\Pow\Dis f$. Hence,
\begin{eqnarray*}
\!\!\!\!\!\!\!\!
C X
\,\,\, &=& \,\,\,
\set{S \subseteq \Dis(X)}{S \neq \emptyset, S = \con(S)}\\
\qquad C f(S)
\,\,\, &=& \,\,\,
\set{\Dis f(\varphi)}{\varphi \in S}.
\end{eqnarray*} 
\end{enumerate}

We will often decompose $\Pow$ as $\Powne+1$ where $\Powne$ is the nonempty powerset functor and $(-)+1$ is the termination functor defined for every set $X$ by $X+1=X\cup \{*\}$ with $*\notin X$ and every function $f\colon X\to Y$ by $f+1(*)=*$ and $f+1(x)=f(x)$ for $x\in X$.
\end{exa}

A category $\cat C$ is concrete, if it admits a canonical forgetful functor $\mathcal U\colon \cat C \to \Sets$. By a forgetful functor we mean a functor that is identity on arrows. Intuitively, a concrete category has objects that are sets with some additional structure, e.g. particular algebras, and morphisms that are a particular kind of functions between the sets, e.g. the corresponding algebra homomorphisms. The forgetful functor forgets the additional structure and maps the objects to the underlying sets. It is identity on arrows in the sense that every such particular kind of function is also just a function between the underlying sets and hence mapped to itself. For example, the category of monoids with monoid homomorphisms as arrows is concrete. The forgetful functor maps every monoid to its carrier set, and every monoid homomorphism to itself (being a function from the one carrier set to the other). 

Let $F\colon \cat C \to \cat D$ and $G\colon \cat C \to \cat D$ be two functors. A natural transformation $\sigma\colon F \Rightarrow G$ is a family of arrows $\sigma_X\colon FX \to GX$ in $\cat D$ such that the following diagram commutes for every arrow $f\colon X \to Y$. 
$$\xymatrix@C=+4pc@R-.5pc{
FX\ar[d]_-{\sigma_X}\ar[r]^-{Ff} & FY\ar[d]^-{\sigma_Y}\\
GX\ar[r]^-{Gf} & GY
}$$

%

\smallskip

 Coalgebras provide an abstract framework for state-based systems. Let $\cat C$ be a base category.
 A coalgebra is a pair $(S,c)$ of a state space $S$ (object in $\cat C$) and an arrow $c\colon S \to FS$ in $\cat C$
  where $F\colon\cat{C} \rightarrow \cat{C}$ is a functor that specifies the type of transitions.

We will sometimes just say the coalgebra $c\colon S \to FS$, meaning the coalgebra $(S,c)$. A coalgebra homomorphism from a coalgebra $(S,c)$ to a coalgebra $(T,d)$ is an arrow $h\colon S \to T$ in $\cat C$ that makes the following diagram commute. 
$$\xymatrix@C=+4pc@R-.5pc{
S\ar[d]_-{c}\ar[r]^-{h} & T\ar[d]^-{d}\\
FS\ar[r]^-{Fh} & FT
}$$

Coalgebras of a functor $F$ and their coalgebra homomorphisms form a category that we denote by $\CoAlg_{\cat C}{(F)}$. 

\begin{exa}\label{ex:coalgebras}
Here are some examples: 
 	\begin{enumerate}[itemsep=1mm]
\item Deterministic automata are coalgebras for the functor $2 \times (-)^L$ for an alphabet $L$.
\item Nondeterministic automata are coalgebras for $2 \times \Pow^L$.
\item Labelled transition systems (LTS) with labels in $L$ are coalgebras for $\Pow^L$.
\item Markov chains are coalgebras for the distribution functor $\Dis$.
\end{enumerate}
\end{exa}

Coalgebras over a concrete category, with $\mathcal U\colon \cat C \to \Sets$ being the forgetful functor, are equipped with a generic behavioural equivalence, which we define next.
Let $(S,c)$ be an $F$-coalgebra on a concrete category $\cat C$. An equivalence relation $R \subseteq \mathcal{U}S \times \mathcal{U}S$ is a kernel bisimulation (synonymously, a cocongruence)~\cite{Staton11,Kurz00:thesis,Wol00:cmcs} if it is the kernel of a homomorphism, i.e., $R = \ker \mathcal{U}h = \{(s,t) \in \mathcal{U}S \times \mathcal{U}S\mid \mathcal{U}h(s) = \mathcal{U}h(t)\}$ for some coalgebra homomorphism $h\colon (S,c) \to (T,d)$ to some $F$-coalgebra $(T,d)$. Two states $s,t$ of a coalgebra are behaviourally equivalent, notation $s \approx t$, if and only if there is a kernel bisimulation $R$ with $(s,t) \in R$. It is not difficult to show that behaviour equivalence is itself a kernel bisimulation. 

Moreover, coalgebras over a concrete category have a \emph{set of states}, namely the set of states of $(S,c)$ is $\mathcal{U}S$. 
The following property is simple but important: if there is a functor from one category of coalgebras (over a concrete category) to another that keeps the set of states the same and is identity on morphisms, then this functor preserves behavioural equivalence, i.e., if two states are equivalent in a coalgebra of the first category, then they are also equivalent in the image under the functor in the second category. 
More precisely, let $\mathcal T\colon \CoAlg_{\cat C}{(F_{\cat C})} \to \CoAlg_{\cat D}{(F_{\cat D})}$ for two concrete categories $\cat C$ and $\cat D$. Assume further that $U_{\cat D} \after \mathcal T = U_{\cat C}$ for $$\xymatrix{{U_{\cat X}\colon \CoAlg_{\cat X}{(F)}} \ar[r]^{\hspace*{1cm}\bar U_{\cat X}} &  \cat X \ar[r]^{\mathcal U_{\cat X}} & \Sets}$$ being the composition of the obvious forgetful functors with $\bar U_{\cat X}(S,c) = S$ and $\mathcal U_{\cat X}$ the forgetful functor making $\cat X$ concrete. Then $\mathcal T$ preserves behavioural equivalence. Note that $U_{\cat D} \after \mathcal T = U_{\cat C}$ on objects means that $\mathcal T$ keeps the state set the same and $U_{\cat D} \after \mathcal T = U_{\cat C}$ on morphisms means that $\mathcal T$ is identity on morphisms. 

We are now in position to connect probabilistic automata to coalgebras.
\begin{propC}[\cite{BSV04:tcs,Sokolova11}]\label{prop:PA-coalg-bis}
	A probabilistic automaton $M = (S, \Act, \to)$ can be identified with a $(\Pow\Dis)^L$-coalgebra $c_M \colon S \to (\Pow\Dis S)^L$ on $\Sets$, where $s \stackrel{a}{\to} \xi$ in $M$ if and only if $\xi \in c_M(s)(a)$ in $(S,c_M)$. Bisimilarity in $M$ equals behavioural equivalence in $(S,c_M)$, i.e., for two states $s,t \in S$ we have $s \sim t \Leftrightarrow s \approx t$. \qed
\end{propC}

\begin{exa}\label{ex:PACoalg}
The PA on the left of Figure \ref{fig:weighted} corresponds to a coalgebra $c_M \colon S \to (\Pow\Dis S)^L$. The set of states is $S=\{x_0,x_1,x_2,x_3,y_0,y_1,y_2,y_3\}$. The transition function $c_M$ is defined as expected. For instance, $c_M(y_0) (a) = \{\delta_{y_3}, \frac{1}{2}\delta_{y_1} + \frac{1}{2}\delta_{y_2}\}$.
\end{exa}

It is also possible to provide convex bisimilarity semantics to probabilistic automata via coalgebraic behavioural equivalence, as the next proposition shows. 
\begin{propC}[\cite{Mio14}]\label{prop:PA-coalg-conv-bis}
	Let $M = (S, \Act, \to)$ be a probabilistic automaton, and let $(S,\bar c_M)$ be a $(C+1)^L$-coalgebra on $\Sets$ defined by $\bar c_M(s)(a) = \con(c_M(s)(a))$ if $c_M(s)(a) = \{ \xi \mid s \stackrel{a}{\to} \xi\} \neq \emptyset$; and $\bar c_M(s)(a) = *$ if $c_M(s)(a) = \emptyset$. Convex bisimilarity in $M$ equals behavioural equivalence in $(S,\bar c_M)$. \qed
\end{propC}

The connection between $(S, c_M)$ and $(S, \bar c_M)$ in Proposition~\ref{prop:PA-coalg-conv-bis} is the same as the connection between $M$ and $M_c$ in Section~\ref{sec:PA}. Abstractly, it can be explained using the following well known generic property.

\begin{lemC}[\cite{Rut00:tcs,BSV04:tcs}]\label{lem:nat-transf}
	Let $\sigma\colon F \Rightarrow G$ be a natural transformation from $F\colon \cat C \to \cat C$ to $G\colon \cat C \to \cat C$. Then $\mathcal T\colon \CoAlg_{\cat C}{(F)} \to \CoAlg_{\cat C}{(G)}$ given by $$\mathcal T(S \stackrel{c}{\to} FS) = (S \stackrel{c}{\to} FS \stackrel{\sigma_S}{\to} GS)$$ on objects and identity on morphisms is a functor that preserves behavioural equivalence. If $\sigma$ is injective, meaning that all components of $\sigma$ are injective, then $\mathcal T$ also reflects behavioural equivalence. \qed
\end{lemC}

\begin{exa}\label{ex:nat-transf-Conv}
	We have that $\con \colon \Pow\Dis \Rightarrow C+1$ given by $\con(\emptyset) = *$ and $\con(X)$ is the already-introduced convex hull for $X \subseteq \Dis S, \, X \neq \emptyset$ is a natural transformation. Therefore, $\con^L \colon (\Pow\Dis)^L \Rightarrow (C+1)^L$ is one as well, defined pointwise. As a consequence from Lemma~\ref{lem:nat-transf}, we get a functor
	$\mathcal T_{\con}\colon \CoAlg_\Sets{((\Pow\Dis)^L)} \to \CoAlg_\Sets{((C+1)^L)}$ and hence bisimilarity implies convex bisimilarity in probabilistic automata. 
	
	On the other hand, we have the injective natural transformation $\iota\colon C + 1 \Rightarrow \Pow\Dis$ given by $\iota(X) = X$ and $\iota(*) = \emptyset$ yielding an injective natural transformation $\chi\colon (C + 1)^L \Rightarrow (\Pow\Dis)^L$. As a consequence, convex bisimilarity coincides with strong bisimilarity on the ``convex-closed'' probabilistic automaton $M_c$, i.e., the coalgebra $(S,\bar c_M)$ whose transitions are all convex combinations of $M$-transitions.
\end{exa}

\begin{exa}\label{ex:PACoalgConv}
The coalgebra $c_M \colon S \to (\Pow\Dis S)^L$ of Example \ref{ex:PACoalg} is mapped by $\mathcal T_{\con}$ into the coalgebra $\bar c_M \colon S \to (C+1)^L$ which is just $\con^L \after c_M $. For instance, $\bar c_M (y_0)(a)= \con \{\delta_{y_3}, \frac{1}{2}\delta_{y_1} + \frac{1}{2}\delta_{y_2}\}$.
\end{exa}

\subsection{Algebras for a Monad}\label{sec:monads}

The behaviour functor $F$ often is, or involves, a monad $\T$,
providing certain computational effects, such as partial,
non-deterministic, or probabilistic computation. 

More precisely, a monad is
a functor $\T\colon\cat{C} \rightarrow \cat{C}$ together with two
natural transformations: a unit $\eta\colon \idmap_{\cat{C}}
\Rightarrow \T$ and multiplication $\mu \colon \T^{2} \Rightarrow
\T$. These are required to make the following diagrams commute, for
$X\in\cat{C}$.
$$\xymatrix@C-.5pc@R-.5pc{
\T X\ar[rr]^-{\eta_{\T X}}\ar@{=}[drr] & & \T^{2}X\ar[d]^{\mu_X} & &
   \T X\ar[ll]_{\T\eta_{X}}\ar@{=}[dll] 
&  & \T^{3}X\ar[rr]^-{\mu_{\T X}}\ar[d]_{\T\mu_{X}} 
   & & \T^{2}X\ar[d]^{\mu_X} \\
& & \T X & &
& &
\T^{2} X\ar[rr]_{\mu_X} & & \T X
}$$

\noindent We briefly describe two examples of monads on $\Sets$:
\begin{itemize}
\item The powerset monad $\Pow$, with unit given by
  singleton $\eta(x) = \{x\}$ and multiplication by union
  $\mu(\setin{X_i}{\Pow X}{i \in I}) = \bigcup_{i \in I} X_i$.
\item The distribution monad $\Dis$, with unit given by a Dirac
distribution $\eta(x) = \delta_x = ( x \mapsto 1)$ for $x \in X$ and
the multiplication by $\mu(\Phi)(x) = \sum\limits_{\varphi \in \supp(\Phi)}
\Phi(\varphi)\cdot \varphi(x)$ for $\Phi \in \Dis\Dis X$.
\end{itemize}

\begin{wrapfigure}{r}{0pt}
\end{wrapfigure}

With a monad $\T$ on a category $\cat{C}$ one associates the  Eilenberg-Moore category
$\EM(\T)$ of Eilenberg-Moore algebras. Objects of
$\EM(\T)$ are pairs $\Alg A = (A, a)$ of an object $A \in \cat C$ and an arrow
$a\colon \T A \rightarrow A$, making the first two
diagrams below commute. 
$$\xymatrix@R-.5pc{
A\ar@{=}[dr]\ar[r]^-{\eta} & \T A\ar[d]^{a}
& 
\T^{2}A\ar[d]_{\mu}\ar[r]^-{\T a} & \T A\ar[d]^{a}
& &
\T A\ar[d]_{a}\ar[r]^-{\T h} & \T B\ar[d]^{b} \\
& A 
&
\T A\ar[r]_-{a} & A
& &
A\ar[r]_-{h} & B
}$$

\noindent A homomorphism from an algebra $\Alg A = (A, a)$ to an algebra $\Alg B = (B, b)$ is a map $h\colon
A\rightarrow B$ in $\cat{C}$ between the underlying objects making the
diagram above on the right commute. The diagram in the middle thus
says that the map $a$ is a homomorphism from $(\T A, \mu_A)$ to $\Alg A$. The
forgetful functor $\mathcal{U}\colon \EM(\T) \rightarrow \cat{C}$ mapping an algebra to its carrier has a
left adjoint $\mathcal{F}$, mapping an object $X\in\cat{C}$ to the
(free) algebra $(\T X, \mu_X)$ generated by $X$. We have that $\T = \mathcal{F}\after\mathcal{U}$. Recall that an algebra $\Alg A = (A, a)$ is a free algebra generated by a set $X \subseteq A$ if any map $f$ from $X$ to the carrier of another algebra $\Alg B = (B, b)$, i.e. $f\colon X \to B$, extends to a unique homomorphism $f^\#$  from $\Alg A$ to $\Alg B$. Free algebras are unique up to isomorphism.

A category of Eilenberg-Moore algebras which is particularly relevant for our exposition is described in the following proposition. See~\cite{swirszcz:1974} and~\cite{semadeni:1973} for the original result, but also~\cite{doberkat:2006,doberkat:2008} or \cite[Theorem~4]{jacobs:2010} where a concrete and simple proof is given.  

\begin{propC}[\cite{swirszcz:1974,doberkat:2006,doberkat:2008,jacobs:2010}]\label{prop:EMD-CA}
Eilenberg-Moore algebras of the finitely supported distribution monad $\Dis$ are exactly convex algebras as defined in Section~\ref{sec:CA}. The arrows in the Eilenberg-Moore category $\EM(\Dis)$ are convex algebra homomorphisms. \qed	
\end{propC}
 
As a consequence, we will interchangeably use the abstract (Eilenberg-Moore algebra) and the concrete definition (convex algebra), whatever is more convenient. For the latter, we also just use binary convex operations, by Proposition~\ref{prop:bin-suff}, whenever more convenient.

\subsection{The Generalised Determinisation}\label{sec:GPC}
We now, briefly and without all details, recall the general determinisation construction from \cite{SBBR10}, which serves as inspiration for our work. Our goal is to understand the belief-state transformer as a particular determinisation of the original PA. Just like determinisation of NDA transforms a nondeterministic automaton $S\to 2\times (\Pow S)^\Act$ into a deterministic automaton on the powerset of original states $\Pow S \to 2\times (\Pow S)^\Act$, we aim at transforming a PA  $S \to (\Pow \Dis S)^{\Act}$ into its belief-state transformer $\Dis S \to (\Pow\Dis S)^{\Act}$.  

A functor $\overline{F}\colon \EM(\T) \to \EM(\T)$ is said to be a lifting of a functor $F\colon \cat C \to \cat C$ if and only if $\mathcal{U} \after \overline{F} = F \after \mathcal{U}$, as shown in the diagram below. Concretely, this implies that $\overline{F}$ maps an algebra with carrier $X$ to an algebra with carrier $FX$.   

$$\xymatrix@C=+4pc@R-.5pc{
{\EM(\T)}\ar[d]_-{\mathcal{U}}\ar[r]^-{\overline{F}} & {\EM(\T)}\ar[d]^-{\mathcal{U}}\\
{\cat C}\ar[r]^-{F} & {\cat C}
}$$
Here, $\mathcal U$ is the 
forgetful functor $\mathcal{U}\colon \EM(\T) \rightarrow \cat{C}$ mapping an algebra to its carrier. It has a
left adjoint $\mathcal{F}$, mapping an object $X\in\cat{C}$ to the
(free) algebra $(\T X, \mu_X)$. We have that $\T =\mathcal{U} \after  \mathcal{F}$. 

Whenever $F\colon \cat C \to \cat C$ has a lifting $\overline{F}\colon \EM(\T) \to \EM(\T)$, one has the following functors between categories of coalgebras. 
$$
\xymatrix@R=.3cm{ &  \CoAlg_{\EM(\T)}{(\overline{F})} \ar[rd]^{\Ubar}\\
 \CoAlg_{\cat C }{(F\T)} \ar[ru]^{\GPC}& & \CoAlg_{\cat C }{(F)}
}
$$

The functor $\GPC$ transforms every coalgebra $c\colon S\to F\T S$ over the base category into a coalgebra $c^\sharp \colon \mathcal{F}S \to \overline{F} \mathcal{F}S$. Note that this is a coalgebra on $\EM(\T)$: the state space carries an algebra, actually the freely generated one, and $c^\sharp$ is a homomorphism of $\T$-algebras. Intuitively, this amounts to compositionality: like in GSOS\footnote{GSOS refers to a general format of  structural operational semantics~\cite{Bloom1995} rules.} specifications, the transitions of a compound state are determined by the transitions of its components.

The functor $\Ubar$ simply forgets about the algebraic structure: $c^\sharp$ is mapped into $$\mathcal{U}c^\sharp \colon \T S = \mathcal{U}\mathcal{F}S \to \mathcal{U} \overline{F} \mathcal{F}S = F \mathcal{U} \mathcal{F} S = F\T S \text{.}$$
An important property of $\Ubar$ is that it preserves and reflects behavioural equivalence. On the one hand, this fact usually allows to give concrete characterisation of  $\approx$ for $\overline{F}$-coalgebras. On the other, it allows, by means of the so-called up-to techniques, to exploit the $\T$-algebraic structure of $\mathcal{F}S$ to check $\approx$ on $\mathcal{U}c^\sharp$.

\medskip

By taking $F = 2\times (-)^\Act$ and $\T = \Pow$, one transforms $c\colon S\to 2\times (\Pow S)^\Act$ into $\mathcal{U}c^\sharp \colon \Pow S \to 2\times (\Pow S)^\Act$. The former is a non-deterministic automaton (every $c$ of this type is a pairing $\langle o,t\rangle$ of $o\colon S\to 2$, defining the final states, and $t\colon S\to \Pow(S)^\Act$, defining the transition relation) and the latter is a deterministic automaton which has $\Pow S$ as states space. In \cite{SBBR10}, see also~\cite{JacobsSS15}, it is shown that, for a certain choice of the lifting $\overline{F}$, this amounts exactly to the standard determinisation from automata theory. This explains why this construction is called \emph{the generalised determinisation}.


In a sense, this is similar to the translation of probabilistic automata into belief-state transformers that we have seen in Section \ref{sec:PA}. Indeed, probabilistic automata are coalgebras $c\colon S \to (\Pow \Dis S)^{\Act}$ and belief state transformers are coalgebras of type $\Dis S \to (\Pow\Dis S)^{\Act}$. One would like to take $F = \Pow^{\Act}$ and $\T = \Dis$ and reuse the above construction but, unfortunately, $\Pow^{\Act}$ does \emph{not} have a \emph{suitable} lifting to $\EM(\Dis)$. This is a consequence of two well known facts: the lack of a \emph{suitable} distributive law $\rho\colon \Dis\Pow \Rightarrow \Pow\Dis$~\cite{VW06}\footnote{As shown in~\cite{VW06}, there is no distributive law of the powerset monad over the distribution monad. Note that a ``trivial'' lifting and a corresponding distributive law of the powerset \emph{functor} over the distribution monad exists, it is based on~\cite{CR11} and has been exploited in~\cite{JacobsSS15}. However, the corresponding ``determinisation'' is trivial, in the sense that its distribution bisimilarity coincides with bisimilarity, and it does not correspond to the belief-state transformer.}
 and the one-to-one correspondence between distributive laws and liftings, see e.g.~\cite{JacobsSS15}. In the next section, we will nevertheless provide a ``powerset-like'' functor on $\EM(\Dis)$ that we will exploit then in Section \ref{sec:Landscape} to properly model PA as belief-state transformers.

\section{Coalgebras on Convex Algebras}\label{sec:IAPAonCA}

In this section we provide several functors on $\EM(\Dis)$ that will be used in the modelling of probabilistic automata as coalgebras over $\EM(\Dis)$. This will make explicit the implicit algebraic structure (convexity)  in probabilistic automata and lead to distribution bisimilarity as natural semantics for probabilistic automata in Section~\ref{sec:Landscape}. Note that we slightly overload the notation here and conveniently use $C$ for denoting convex subsets of a convex algebra, hoping that this does not bring confusion with the nonempty convex subsets of distributions functor which is also denoted by $C$.

\subsection{Convex Powerset on Convex Algebras}\label{sec:CAPower}

We now define a functor, the (nonempty) convex powerset functor, on $\EM(\Dis)$. Let $\Alg A$ be a convex algebra. We define $\PowC\Alg A$ to be $\PowC\Alg A = \Alg A_c = (A_c, a_c)$ where 
 $$A_c = \{ C \subseteq A \mid C \neq \emptyset,\,C \text{ is convex}\}$$ and $a_c$ is the convex algebra structure given by the following \emph{pointwise} binary convex combinations:
 $$pC + \bar pD = \{pc + \bar p d \mid c \in C, d \in D\}.$$
It is important that we only allow nonempty convex subsets in the carrier $A_c$ of $\PowC\Alg A$, as otherwise the projection axiom fails. 

\begin{exa}
In Example~\ref{ex:convexalgebra}, we have discussed the convex algebra $\Dis(X)$ for $X=\{x,y,z\}$ and we have given several examples of convex subsets of  $\Dis(X)$. All those convex subsets belong to $\PowC\Dis(X)$ apart from $\emptyset$. Observe that, according to the definition above $p C + \bar p \emptyset$ would be equal to $\emptyset$ violating the projection axiom (as noted above). 

Consider again the sets $C_1$ and $C_2$ from Example~\ref{ex:convexalgebra} and their convex closures:
$$\begin{array}{rclcl}
\con(C_1) &=& \con\{\delta_x, \delta_y\} &=& \set{\varphi\in \Dis(X)}{\varphi(z)=0}\\
\con(C_2) &=&  \con \{\delta_x, \frac{1}{2}\delta_x+ \frac{1}{2}\delta_y\} &=& \set{\varphi\in\Dis(X)}{\varphi(x)\geq \frac{1}{2},\; \varphi(y)=1-\varphi(x)}\text{.}\end{array}$$
By Definition $\frac{1}{2}\con(C_1) + \frac{1}{2} \con(C_2)$ is the convex subset $\set{\frac{1}{2} \varphi_1 + \frac{1}{2} \varphi_2}{\varphi_i \in \con(C_i)}$ which, as we will demonstrate, equals the set $\set{\varphi\in \Dis(X)}{\varphi(x)\geq \frac{1}{4},\; \varphi(y)=1-\varphi(x)}$. To start with, it is important to notice that $$\frac{1}{2}\con(C_1) + \frac{1}{2} \con(C_2) = \con\left(\frac{1}{2}{C_1} + \frac{1}{2}{C_2}\right)$$ which can be shown for the convex algebra on $\Dis(X)$. Hence, we focus on $\frac{1}{2} \varphi_1 + \frac{1}{2} \varphi_2$ for $\varphi_1$ and $\varphi_2$ in the generating sets $C_1 = \{\delta_x, \delta_y\}$ and $C_2 = \{\delta_x, \frac{1}{2}\delta_x+ \frac{1}{2}\delta_y\}$, respectively. We obtain the following four distributions:
$$
\begin{array}{rcl}
\frac{1}{2}\delta_x +\frac{1}{2} \delta_x &=& \delta_x \\[2mm]
\frac{1}{2} \delta_x + \frac{1}{4}\delta_x+ \frac{1}{4}\delta_y & = &\frac{3}{4}\delta_x + \frac{1}{4}\delta_y\\[2mm]
\frac{1}{2}\delta_y + \frac{1}{2} \delta_x & = & \frac{1}{2}\delta_x + \frac{1}{2} \delta_y  \\[2mm]
\frac{1}{2} \delta_y + \frac{1}{4}\delta_x+ \frac{1}{4}\delta_y &= & \frac{1}{4} \delta_x + \frac{3}{4} \delta_y
\end{array}
$$

\noindent By noting that both $\frac{3}{4}\delta_x + \frac{1}{4}\delta_y$ and $\frac{1}{2}\delta_x + \frac{1}{2} \delta_y$  belong to $\con \{ \delta_x, \frac{1}{4} \delta_x + \frac{3}{4} \delta_y\}$ one can readily see that $\con\{ \delta_x , \,\frac{3}{4}\delta_x + \frac{1}{4}\delta_y, \, \frac{1}{2}\delta_x + \frac{1}{2} \delta_y, \,\frac{1}{4} \delta_x + \frac{3}{4} \delta_y \} = \con \{ \delta_x, \frac{1}{4} \delta_x + \frac{3}{4} \delta_y\}$. Finally, it is easy to show that the latter is the set $\set{\varphi\in\Dis(X)}{\varphi(x)\geq \frac{1}{4},\; \varphi(y)=1-\varphi(x)}$.

\end{exa}

\begin{rem}
For convex subsets of a finite dimensional vector space, the pointwise operations are known as 
		the Minkowski addition and are a basic construction in convex geometry, see e.g.~\cite{schneider:1993}. 
The pointwise way of defining algebras over subsets (carriers of subalgebras) has also been studied in universal algebra, see e.g.~\cite{BosnjakM07,BosnjakM03,Brink93}. 
\end{rem}

Next, we define $\PowC$ on arrows of $\EM(\Dis)$. For a convex homomorphism $h\colon \Alg A \to \Alg B$, we fix $\PowC h = \Pow h$.
The following property ensures that we are on the right track, namely that $\PowC\colon \EM(\Dis) \to \EM(\Dis)$ is a functor. 

\begin{prop}\label{prop:Pc-functor}
	
	$\PowC\Alg A$ is a convex algebra. If $h \colon \Alg A \to \Alg B$ is a convex homomorphism, then so is $\PowC h\colon \PowC\Alg A \to \PowC\Alg B$. $\PowC$ is a functor on $\EM(\Dis)$.
\end{prop}

\begin{proof}
 Due to Proposition~\ref{prop:bin-suff}, to prove that $\PowC\Alg A$ is a convex algebra, all we need is to check (1) idempotence, (2) parametric commutativity, and (3) parametric associativity. 	
 \begin{enumerate}
	\item $C \subseteq pC + \bar pC$ as $c = pc + \bar p c \in pC + \bar pC$. For the opposite inclusion, consider $pc_1 + \bar pc_2 \in pC + \bar p C$ for $c_1, c_2 \in C$. As $C$ is convex, $pc_1 + \bar pc_2 \in C$. Hence idempotence holds. 
	\item Follows from parametric commutativity in $\Alg A$. We have $$pC + \bar pD = \{pc + \bar p d \mid c \in C, d\in D\} = 
	\{\bar pd + p c \mid d \in D, c \in C\} = \bar p D + p C$$ proving parametric commutativity. 
	\item Similarly, parametric associativity follows from parametric associativity in $\Alg A$: 
	\begin{eqnarray*} p(qC + \bar q D) + \bar p E & = & 
	\{p(qc + \bar q d)+\bar p e \mid c \in C, d \in D, e \in E\}\\
	& = &
	\{pqc + \overline{pq}\left(\frac{p\bar q}{\overline{pq}} d + \frac{\bar p}{\overline{pq}} e\right) \mid c \in C, d \in D, e \in E\}\\
	& = &
	pqC + \overline{pq}\left(\frac{p\bar q}{\overline{pq}}D + \frac{\bar p}{\overline{pq}} E\right).
	\end{eqnarray*}
 \end{enumerate}

Furthermore, we have, straightforwardly, 
	\begin{eqnarray*}
		\PowC h (p C + \bar p D) & = & h (p C + \bar p D)\\
		& = & h (\{ pc + \bar p d \mid c \in C, d \in D\})\\
		& = & \{ h(pc + \bar p d) \mid c \in C, d \in D\}\\
		& = & \{p h(c) + \bar p h(d) \mid c \in C, d \in D\}\\
		& = & p h(C) + \bar p h(D)\\
		& = & p \PowC h(C) + \bar p \PowC h(D). 
	\end{eqnarray*}

Proposition~\ref{prop:EMD-CA} now implies the property. 
\end{proof}

\begin{rem}\label{lem:Pc-liftings}
	$\PowC$ is not a lifting of the nonempty convex subsets of distributions functor  $C$ to $\EM(\Dis)$, yet it satisfies an equation close to a lifting, namely $C= \mathcal{U}\after \PowC \after \mathcal{F}$ as illustrated below on the left. Moreover, $\PowC$ is also not a lifting of $\Powne$, the nonempty powerset functor, but there exists a natural embedding $e\colon \mathcal U \after \PowC \Rightarrow \Powne \after \mathcal U$ given by $e(C) = C$, i.e., we are in the situation:
	  $$\qquad\qquad\qquad\xymatrix@R-1.8pc@C-1pc{
\EM(\Dis)\ar[rr]^-{\PowC} & & \EM(\Dis)\ar[dd]^{\mathcal{U}} &  \qquad\qquad&
\EM(\Dis)\ar[rr]^-{\PowC}\ar[dd]_{\mathcal{U}} & & \EM(\Dis)\ar[dd]^{\mathcal{U}}\\ & & & & & \supseteq &\\
\Sets\ar[rr]^-{C} \ar[uu]^{\mathcal{F}} & & \Sets & &
\Sets\ar[rr]^-{\Powne}  & & \Sets &&  
}$$
\end{rem}

The right diagram in Remark~\ref{lem:Pc-liftings} simply states that every convex subset is a subset, but this fact and the natural transformation $e$ are useful in the sequel. In particular, using $e$ we can show the next result.

\begin{prop}\label{prop:Pc-monad}
	$\PowC$ is a monad on $\EM(\Dis)$, with $\eta$ and $\mu$ as for the powerset monad.
\end{prop}

\begin{proof}
	Let $\Alg X$ be a convex algebra and consider $\PowC\Alg X$.
	We have $\eta(x) = \{x\}$ is a convex subset, as every singleton is. Moreover,  $\eta$ is a convex homomorphism as $p\{x\} + \bar p \{y\} = \{px + \bar py\}$. We have $\eta$ (of $\PowC$) is natural if and only if the upper square of the left diagram below commutes. 
		$$\xymatrix@R-1.5pc@C-1pc{
{\mathcal U\Alg X}\ar[dd]^{\eta_{\Alg X}}\ar[rr]^-{\mathcal U f}\ar@/_5ex/[dddd]_{\eta_X} & & {\mathcal U\Alg Y}\ar[dd]_{\eta_{\Alg Y}}\ar@/^5ex/[dddd]^{\eta_Y} & & &&
{\mathcal U\PowC\PowC\Alg X}\ar[dd]^{\mu_{\Alg X}}\ar[rr]^-{\mathcal U\PowC\PowC f}\ar@/_5ex/[dddd]_{\mu_X} & & {\mathcal U\PowC\PowC\Alg Y}\ar[dd]_{\mu_{\Alg Y}}\ar@/^5ex/[dddd]^{\mu_Y}\\
&&&&&&&&\\
{\mathcal U\PowC\Alg X}\ar[dd]^{e_{\Alg X}}\ar[rr]^-{\mathcal U\PowC f} & & {\mathcal U\PowC\Alg Y}\ar[dd]_{e_{\Alg Y}} & & &&
{\mathcal U\PowC\Alg X}\ar[dd]^{e_{\Alg X}}\ar[rr]^-{\mathcal U\PowC f} & & {\mathcal U\PowC\Alg Y}\ar[dd]_{e_{\Alg Y}}\\
&&&&&&&&\\
{\Pow\mathcal U\Alg X}\ar[rr]^-{\Pow\mathcal U f} & & {\Pow\mathcal U\Alg Y} & & &&
{\Pow\mathcal U\Alg X}\ar[rr]^-{\Pow\mathcal U f} & & {\Pow\mathcal U\Alg Y}
}$$
	However, the outer square of the diagram does commute - due to naturality of $\eta$ (of $\Pow$), the lower square does commute - due to naturality of $e$, the outside triangles also do - due to the definitions of both $\eta$'s and $e$, and $e$ is injective. As a consequence, the upper square commutes as well. 
	
	For $\mu$, notice that also $\mu_{\Alg X}$ is a convex homomorphism from $\PowC\PowC\Alg X$ to $\PowC\Alg X$, and all the arguments that we used for naturality of $\eta$ apply to the naturality of $\mu$ (of $\PowC$) as well, when looking at the right diagram above. So, $\mu$ is natural as well. 
	
	Clearly, $\eta$ and $\mu$ (of $\PowC$) satisfy the compatibility conditions of the definition of a monad, as so do $\eta$ and $\mu$ (of $\Pow$). 
\end{proof}

\subsection{Termination on Convex Algebras}\label{sec:Lift}

The functor $\PowC$ defined in the previous section allows only for nonempty convex subsets. We still miss a way to express termination. 
The question of termination amounts to the question of extending a convex algebra $\Alg A$ with a single element $*$. This question turns out to be rather involved, beyond the scope of this paper. 
 The answer from~\cite{SW18} is: there are many ways to extend any convex algebra $\Alg A$ with a single element, but there is only one natural functorial way. 
 Somehow now mathematics is forcing us the choice of a specific computational behaviour for termination! 

Given a convex algebra $\Alg A$, let $\Alg A + 1$ have the carrier $A + \{*\}$ for $* \notin A$ and convex operations for $p \in (0,1)$ given by 
\begin{equation}\label{eq:black-hole}
			px \oplus \bar py = 
			\begin{cases}
				px + \bar p y &\hspace*{-3mm},\quad x, y \in A,\\
				* &\hspace*{-3mm},\quad x = * \text{ or } y = *.
			\end{cases}
		\end{equation}
Here, the newly added $*$ behaves as a black hole that attracts every other element of the algebra in a convex combination. It is worth to remark that this extension is folklore~\cite{fritz:2015}. 

\begin{propC}[\cite{SW18,fritz:2015}]\label{prop:Lift-on-CA}
	$\Alg A +1$ as defined above is a convex algebra that extends $\Alg A$ by a single element.  The map $h+1$ obtained with the termination functor in $\Sets$ is a convex homomorphism if $h\colon \Alg A \to \Alg B$ is. The assignments $(-)+1$ give a functor on $\EM(\Dis)$. \qed
\end{propC}

We call the functor $(-)+1$ on $\EM(\Dis)$ the termination functor, due to the following property that follows directly from the definitions.  

\begin{lem}\label{lem:lift-lifting}
	The functor $(-)+1$ is a lifting of the termination functor to  $\EM(\Dis)$.  \qed
\end{lem}

\begin{rem}\label{rem:no-coproduct}
	Note that we are abusing notation here: our termination functor $(-) +1$ on $\EM(\Dis)$ is not the coproduct $(-) + 1$ in $\EM(\Dis)$.  
	The coproduct was concretely described in \cite[Lemma~4]{jacobs.westerbaan.westerbaan:2015}, and the coproduct $\Alg X + 1$ 
	has a much larger carrier than $X  + 1$. Nevertheless, we use the same notation as it is very intuitive and due to Lemma~\ref{lem:lift-lifting}. 
\end{rem}

\subsection{Constant Exponent on Convex Algebras}\label{sec:ExpA}

We now show the existence of a constant exponent functor on $\EM(\Dis)$. Let $L$ be a set of labels or actions. Let $\Alg A$ be a convex algebra. Consider $\Alg A^L$ with carrier $A^L = \{f \mid f\colon L \to A\}$ and operations defined (pointwise) by $(pf + \bar p g)(l) = pf(l) + \bar pg(l)$.
 
The following property follows directly from the definitions.

\begin{prop}\label{prop:Exp-functor-on-CA}
	$\Alg A^L$ is a convex algebra. If $h \colon \Alg A \to \Alg B$ is a convex homomorphism, then so is $h^L\colon \Alg A^L \to \Alg B^L$ defined as in $\Sets$. Hence, $(-)^L$ defined above is a functor on $\EM(\Dis)$.  \qed
\end{prop}

We call $(-)^L$ the constant exponent functor on $\EM(\Dis)$. The name and the notation is justified by the following (obvious) property.

\begin{lem}\label{lem:Exp-lifting}
	The constant exponent $(-)^L$ on $\EM(\Dis)$ is a lifting of the constant exponent functor $(-)^L$ on $\Sets$. \qed
\end{lem}

\begin{exa}\label{ex:Pc-on-free} Consider a free algebra $\mathcal{F}S = (\Dis S, \mu)$ of distributions over the set $S$.
By applying first the functor $\PowC$, then $(-)+1$ and then $(-)^L$, one obtains the algebra
$$\Definitive{\mathcal{F}S} = \ensuremath{\left(\xy
(0,4)*{\Dis \left((CS+1)^L\right)};
(0,-4)*{(CS+1)^L};
{\ar^{\alpha} (0,2); (0,-2)};
\endxy\right)}$$ where
$CS$ is the set of non-empty convex subsets of distributions over $S$,
and $\alpha$ corresponds to the convex operations\footnote{In this case, for future reference, it is convenient to spell out the $n$-ary convex operations.} $\sum p_i f_i$ defined by
\begin{equation*}
	\left( \sum p_i f_i \right) (l)   = \left\{ \begin{array}{ll}
 	\left\{ \sum p_i \xi_i  \mid \xi_i \in f_i(l)\right\} &  f_i(l) \in  CS \text{~for~all~} \, i \in \{ 1, \dots, n\}\\
 	\,\,\,* & f_i(l) = *  \text{~for~some~} i \in \{ 1, \dots, n\}
 \end{array}
 \right.
\end{equation*}

\end{exa}

\subsection{Transition Systems on Convex Algebras}\label{ssec:TSCA}

We now compose the three functors introduced above to properly model transition systems as coalgebras on $\EM(\Dis)$.
The functor that we are interested in is $\Definit\colon \EM(\Dis) \to \EM(\Dis)$. A coalgebra $(\Alg{S},c)$ for this functor can be thought of as a transition system with labels in $L$ where the state space carries a convex algebra and the transition function $c\colon \Alg{S} \to \Definitive{\Alg{S}}$ is a homomorphism of convex algebras. This property entails compositionality: the transitions of a composite state $\psum{p}{x_1}{x_2}$ are fully determined by the transitions of its components $x_1$ and $x_2$, as shown in the next proposition. We write $x \Tr{a} y$ for $x, y \in S$, the carrier of $\Alg S$ if $y \in c(x)(a)$, and $x \not \Tr{a}$ if $c(x)(a) = *$.

\begin{prop}\label{prop:key}
Let $({\Alg S},c)$ be a  $\Definit$-coalgebra, and let $x_1,x_2,y_1,y_2,z$ be elements of $S$, the carrier of ${\Alg S}$. Then, for all $p\in (0,1)$ and $a\in L$\begin{itemize}[itemsep=1mm]
\item $\psum{p}{x_1}{x_2} \Tr{a} z $ if and only if $z = \psum{p}{y_1}{y_2}$, $x_1\Tr{a}y_1$ and $x_2\Tr{a}y_2$;
\item $\psum{p}{x_1}{x_2} \not \Tr{a}$ if and only if $x_1 \not \Tr{a}y_1$ or $x_2 \not \Tr{a}y_2$. 
\end{itemize}
\end{prop}

\begin{proof}
Since $c$ is a convex algebra homomorphism, we have that for all $p\in (0,1)$ and $a\in L$, $c(\psum{p}{x_1}{x_2})(a) = (\psum{p}{c(x_1)}{c(x_2})) (a)$. The latter is equal, by definition of $(-)^L$ (see Section \ref{sec:ExpA}), to  $\psum{p}{c(x_1)(a)}{c(x_2})(a)$. If there is $i \in \{1,2\}$ such that $c(x_i)(a) = *$, then $\psum{p}{c(x_1)(a)}{c(x_2})(a) = *$ (see Section \ref{sec:Lift}). If not, then both $c(x_1)(a)$ and $c(x_2)(a)$ are in $\PowC \Alg{S}$: $\psum{p}{c(x_1)(a)}{c(x_2})(a)$ is by definition (see Section \ref{sec:CAPower}) the set $\{ \psum{p}{y_1}{y_2} \mid y_1\in c(x_1)(a) \text{ and } y_2\in c(x_2)(a) \}$.
\end{proof}

\begin{exa}\label{ex:alg-transitions}
	Consider the PA in Figure \ref{fig:weighted} (left) and the belief-state transformer generated by it (right). 
	We have $\frac{1}{2}y_1+\frac{1}{2}y_2 \Btr{a} \frac{1}{4}y_1+\frac{3}{4}y_2$ since $y_1 \Btr{a} \frac{1}{2}y_1+\frac{1}{2}y_2$ and $y_2 \Btr{a} y_2$ and $$\frac{1}{2}\left( \frac{1}{2}y_1+\frac{1}{2}y_2\right)+\frac{1}{2}y_2 =  \frac{1}{4}y_1+\frac{3}{4}y_2.$$ Every transition of $\frac{1}{2}y_1+\frac{1}{2}y_2$ arises in this way.
\end{exa}

Transition systems on convex algebras are the bridge between PA and LTSs. In the next section we will show that one can transform an arbitrary PA into a $\Definit$-coalgebra and that, in the latter, behavioural equivalence coincides with the standard notion of bisimilarity for LTSs (Proposition \ref{prop:congruence}).

\section{From PA to Belief-State Transformers}\label{sec:Landscape}

We will now focus on explaining the genesis of the belief-state transformer, as announced in Section~\ref{sec:GPC}.
Recall from Remark~\ref{lem:Pc-liftings} how $\PowC$ is related to $C$ and $\Powne$, namely $C= \mathcal{U}\after \PowC \after \mathcal{F}$ and there is an embedding natural transformation $e\colon \mathcal U \after \PowC \Rightarrow \Powne \after \mathcal U$. The following definition is the obvious generalisation.
It is convenient for us to make this further step of abstraction in order to understand the big picture and avoid repetition and ad-hoc solutions. Our main observation is that in a situation like in the following definition, one can perform generalised determinisation without having a lifting. 

\begin{defi}\label{def:L1L2}
Let $\T \colon \Sets \to \Sets$ be a monad with $\T = \mathcal U \after \mathcal F$ and $\Lone, \Ltwo \colon \Sets \to \Sets$ be two functors. A functor $\High \colon \EM(\T) \to \EM(\T)$ is 
\begin{itemize}
	\item a \emph{quasi lifting} of $\Lone$ if the diagram on the left commutes.
	\item a \emph{lax lifting} of $\Ltwo$ if there exists an  injective natural transformation $e\colon \mathcal{U} \circ \High \Rightarrow \Ltwo \circ \mathcal{U}$ as depicted on the right.
	\item an \emph{$(\Lone,\Ltwo)$ quasi-lax lifting} if it is both a quasi lifting of $\Lone$ and a lax lifting of $\Ltwo$.
\end{itemize}
$$\qquad\qquad\qquad\xymatrix@R-1.8pc@C-1pc{
\EM(\T)\ar[rr]^-{\High} & & \EM(\T)\ar[dd]^{\mathcal{U}} & & 
\EM(\T)\ar[rr]^-{\High}\ar[dd]_{\mathcal{U}} & & \EM(\T)\ar[dd]^{\mathcal{U}} \ar@{}[ddll]^{}\\ & & &  & & \rotatebox{45}{$\Leftarrow$}  &\\
\Sets\ar[rr]^-{\Lone} \ar[uu]^{\mathcal{F}} & & \Sets & &
\Sets\ar[rr]^-{\Ltwo}  & & \Sets &&\qquad\quad \qedd
}$$ 
\end{defi}
So, for instance, $\PowC$ is a $(C,\Powne)$ quasi-lax lifting by Remark~\ref{lem:Pc-liftings}. From this fact, it follows that $\Definit{}$ is a $((C+1)^L,(\Powne+1)^L)$ quasi-lax lifting.
Another interesting example is the generalised determinisation (Section \ref{sec:GPC}): it is easy to see that $\overline{F}$ is a $(F\T,F)$-quasi-lax lifting. 
Like in the generalised powerset construction, in the presence of a quasi-lax lifting one can construct the following functors.
$$
\xymatrix@R=.3cm{ &  \CoAlg_{\EM(\T)}{(\High)} \ar[rd]^{\Ubar}\\
 \CoAlg_{\Sets}{(\Lone)} \ar[ru]^{\GPC}& & \CoAlg_{\Sets}{(\Ltwo)}
}
$$
We first define $\GPC$. Take an $\Lone$-coalgebra $(S,c\colon S \to \Lone S)$ and recall that $\mathcal{F}S$ is the free algebra $\mu \colon \T \T S \to \T S$. The left diagram in Definition \ref{def:L1L2} entails that $\High\mathcal F S$ has as carrier set  $\Lone S$, so  $\High\mathcal F S$  is an algebra $\alpha\colon \T \Lone S \to \Lone S$. We call $\mathcal{U}c^\sharp$ the composition $\mathcal{U}\mathcal{F}S =  \T S \stackrel{\T c}{\longrightarrow}\T\Lone S \stackrel{\alpha}{\longrightarrow}\Lone S = \mathcal{U}\High \mathcal{F} S$. The next lemma shows that $c^\sharp \colon \mathcal{F}S  \to \High \mathcal{F} S$ is a map in $\EM(\T)$.

\begin{lem}\label{lem:GPC-quasi}
There is a 1-1 correspondence between $\Lone$-coalgebras on $\Sets$ and $\High$-coalgebras on $\EM(\T)$ with carriers free algebras, if $\High$ is a quasi lifting of $\Lone$:
$$\begin{prooftree}
{c\colon S \to \Lone S \,\,\,\text{~in~}\,\,\, \Sets}
\Justifies
{c^\#\colon \mathcal{F}S \to \High \mathcal{F} S \,\,\,\text{~in~}\,\,\, \EM(\T)}	
\end{prooftree}$$
\begin{itemize}
\item given $c$, we have $\mathcal{U}c^\# = \alpha\after \T c$ for $\alpha = \High\mathcal{F}S$,
\item given $c^\#$, we have $c = \mathcal{U}c^\# \after \eta$.
\end{itemize}
The assignment $\GPC(S,c) = (\mathcal F S, c^\#)$ and $\GPC(h) = \T h$ gives a functor $\GPC\colon \CoAlg_{\Sets}{(\Lone)} \to \CoAlg_{\EM(\T)}{(\High)}$. 
\end{lem}

\begin{proof}
First, given $c$, consider the map $\alpha\after \T c$. We need to show that $\alpha\after \T c$ is an algebra homomorphism from the free algebra $\mathcal{F}S$ to $H\mathcal{F} S$ in $\EM(\T)$. This will show that $c^\# \colon  \mathcal{F}S \to H\mathcal{F} S$ and $\mathcal{U}c^\# = \alpha\after \T c$. 

Note that $$\ensuremath{\left(\xy
(0,4)*{{\T\Lone S}};
(0,-4)*{{\Lone S}};
{\ar^{\alpha} (0,2); (0,-2)};
\endxy\right)}
 = \High \ensuremath{\left(\xy
(0,4)*{{\T\T S}};
(0,-4)*{{\T S}};
{\ar^{\mu} (0,2); (0,-2)};
\endxy\right)}
.$$

The needed homomorphism property holds since the following diagram commutes:
$$\xymatrix{
\T\T S \ar[d]_{\mu} \ar[rr]^{\T\T c}   &&   \T \T \Lone S \ar[rr]^{\T\alpha} \ar[d]^{\mu}  && \T \Lone S \ar[d]^{\alpha}\\
\T S  \ar[rr]^{\T c}    &&   \T \Lone S \ar[rr]^{\alpha}  && \Lone S \\
}
$$
as the left square commutes by the naturality of $\mu$ and the right one by the Eilenberg-Moore law for $\alpha$. 

Next, we show that the assignments $c \stackrel{\triangleright}{\mapsto} c^
\#$ and $c^\# \stackrel{\triangleleft}{\mapsto} c$ are inverse to each other.

We have $\triangleleft \after \triangleright (c) = c$ as $\alpha \after \T c \after \eta \stackrel{\text{nat.} \eta}{=} \alpha \after \eta \after c \stackrel{\EM-\text{law}}{=} c$. 
Also, we have $\triangleright \after \triangleleft (c^\#) = c^\#$ as
\begin{eqnarray*}
	\alpha \after \T (\mathcal U c^\# \after \eta) & = & \alpha \after \T\mathcal{U} c^\# \after \T\eta\\
	& \stackrel{(*)}{=} & \mathcal U c^\#\after \mu \after \T\eta\\
	& \stackrel{(**)}{=} & \mathcal{U} c^\#
\end{eqnarray*}
where the equality marked by $(*)$ holds since $\mathcal U c^\#$ is an algebra homomorphism, proven above, and the equality marked by $(**)$ holds by the monad law.

By the above, $\GPC$ is well defined on objects. It remains to prove that for two $\Lone$-coalgebras on $\Sets$ $(S, c_S)$ and $(T, c_T)$, and a coalgebra homomorphism $h\colon (S,c_S) \to (T,c_T)$ with $c_T \after h = \Lone h \after c_S$ we have that $\T h$ is a coalgebra homomorphism in $\EM(\T)$ from $(\mathcal{F} S, c_S^\#)$ to $(\mathcal{F} T, c_T^\#)$.

We have 
$$\xymatrix{
\mathcal{U}\mathcal{F} S  \ar[rr]^{\mathcal{U}\mathcal{F} h} \ar[d]_{\mathcal{U}\mathcal{F} c_S} \ar@/_9ex/[dd]_{\mathcal{U}c_S^\#} && \mathcal{U}\mathcal{F} T \ar[d]^{\mathcal{U}\mathcal{F} c_T} \ar@/^9ex/[dd]^{\mathcal{U}c_T^\#}\\
\mathcal{U}\mathcal{F} \Lone S \ar[rr]^{\mathcal{U}\mathcal{F} \Lone h} \ar[d]_{\alpha_S} && \mathcal{U}\mathcal{F} \Lone T \ar[d]^{\alpha_T} \\
\mathcal{U}\High\mathcal{F} S \ar[rr]^{\Lone h} && \mathcal{U}\High\mathcal{F} T
}$$
where the outer triangles commute by definition; the upper square commutes by assumption, i.e., since $h$ is a homomorphism and $\mathcal{U}$ and $\mathcal{F}$ are functors; and the lower square simply states that $\High\mathcal{F} h$ is an arrow in $\EM(\T)$ which of course holds as $\High$ and $\mathcal F$ are functors. 
\end{proof}

\begin{exa}\label{ex:FREEPA}
In Example \ref{ex:PACoalgConv}, we have illustrated the coalgebra $\bar c_M \colon S \to (C+1)^L$ corresponding to the PA in Figure \ref{fig:weighted}.
The functor $\GPC$ maps it into a $(\PowC+1)^L$-coalgebra on $\EM(\Dis)$. The state space is the convex algebra freely generated by $S$, that is $(\Dis(S),\mu)$.
The transition function $\mathcal{U} \bar c_M^\# \colon  \Dis(S) \to \mathcal{U}(\PowC \Dis(S)+1)^L$ is defined as $\alpha \circ \Dis(\bar c_M)$ where $\alpha$ is the algebra from Example \ref{ex:Pc-on-free}.
Transitions are like in Example \ref{ex:alg-transitions}.
\end{exa}

Now we can define $\Ubar \colon \CoAlg_{\EM(\T)}{(\High)}   \to   \CoAlg_{\Sets}{(\Ltwo)}$ as mapping every coalgebra $(\Alg S,c)$ with $c\colon \Alg{S} \to \High \Alg{S}$ into 
\begin{equation*}
\Ubar(\Alg S, c) = (\mathcal{U}\Alg{S}, e_{\Alg S} \after \mathcal{U} c) \text{~~where~~} \mathcal{U}\Alg{S} \stackrel{\mathcal{U}c}{\longrightarrow}   \mathcal{U}\High \Alg{S} \stackrel{e_{\Alg{S}}}{\longrightarrow}   \Ltwo \mathcal{U} \Alg{S}
\end{equation*}
and every coalgebra homomorphism $h\colon (\Alg{S},c) \to (\Alg{T},d)$ into $\Ubar h = \mathcal{U}h$. Routine computations confirm that $\Ubar$ is a functor.

It is immediate to see that $\Ubar \colon \CoAlg_{\EM(\T)}{(\High)}   \to   \CoAlg_{\Sets}{(\Ltwo)}$ preserves behavioral equivalence: if two states of a coalgebra $(\Alg{S},c)$ in $\CoAlg_{\EM(\T)}{(\High)}$ are behaviourally equivalent, then they are also equivalent on $\Ubar(\Alg{S},c)$. Indeed, since $\Ubar$ is a functor that keeps the state set constant and is identity on morphisms, every kernel bisimulation on $(\Alg{S},c)$ is also a kernel bisimulation on $\Ubar(\Alg{S},c)$. The converse is not true in general, as illustrated below.

\begin{exa}
For sake of simplicity we now consider a different PA. Let us take the set of labels $L$ to be $\{a\}$; the set of states $S$ to be $\{x,y\}$; as transitions we only have: $x\stackrel{a}{\to}\delta_x$ and  $y\stackrel{a}{\to}\delta_y$. By applying the construction that we have seen so far, we obtain a $\Pow^L$-coalgebra on $\Sets$, call it $c$, where the state space is $\Dis(S)$ and transitions are given as $\zeta \Btr{a} \zeta$ for all $\zeta \in \Dis(S)$. Consider now the function $h\colon \Dis(S) \to \Dis(S)$ behaving like the identity except on $\delta_y$ which is mapped to $\delta_x$. It is immediate to see that $h\colon (\Dis(S),c) \to (\Dis(S),c)$ is a coalgebra homomorphism and thus the kernel of $h$, namely the relation $R=\set{(\zeta,\zeta)}{\zeta\in \Dis{X}} \cup \{(\delta_x,\delta_y), \; (\delta_y,\delta_x)\}$, is a kernel bisimulation on $\Sets$.
However the function $h$ is not an algebra homomorphism from $(\Dis(S),\mu)$ to  $(\Dis(S),\mu)$ since $h(\frac{1}{2}\delta_x+\frac{1}{2}\delta_y) = \frac{1}{2}\delta_x+\frac{1}{2}\delta_y \neq \delta_x = \frac{1}{2}h(\delta_x)+\frac{1}{2}h(\delta_y)$. Moreover, it is easy to see that any other map with kernel $R$ is not an algebra homomorphism. Therefore, $R$ cannot be a kernel bisimulation on $\EM(\Dis)$.
\end{exa}

We can however state a precise correspondence: a kernel bisimulation $R$ on $\Ubar(\Alg{S},c)$ is a kernel bisimulation on $(\Alg{S},c)$ only if it is a \emph{congruence} with respect to the algebraic structure of $\Alg{S}$. Formally, $R$ is a congruence if and only if the set $\mathcal{U}\Alg{S} / R$ of equivalence classes of $R$ carries an Eilenberg-Moore algebra and the function $ \mathcal{U} [-]_R \colon \mathcal{U}\Alg{S} \to \mathcal{U}\Alg{S} / R$ mapping every element of $\mathcal{U}\Alg{S}$ to its $R$-equivalence class is an algebra homomorphism.

\begin{prop}\label{prop:epi}
The following are equivalent:
\begin{itemize}[itemsep=1mm]
\item $R$ is a kernel bisimulation on $(\Alg{S},c)$,
\item $R$ is a congruence of $\Alg{S}$ and a kernel bisimulation of $\Ubar(\Alg{S},c)$.
\end{itemize} 
\end{prop}

\begin{proof}
We already discussed above that if $R$ is a kernel bisimulation on $(\Alg{S},c)$, then it is a kernel bisimulation of $\Ubar(\Alg{S},c)$.
Moreover, any kernel bisimulation of a coalgebra carried by an algebra is a congruence, as kernels of algebra homomorphisms are congruences.

For the other implication, assume that $R$ is a congruence and a kernel bisimulation of $\Ubar(\Alg{S},c)$. Then $\mathcal{U}\Alg{S}/R$ carries an $\T$-algebra, denoted by $\Alg{S}/R$, and $[-]_R \colon \Alg{S}\to \Alg{S}/R$ is a map in $\EM(\T)$. If $R$ is a kernel bisimulation on $\Ubar(\Alg{S},c)$, then there exists a function $f\colon \mathcal{U}(\Alg{S}/ R) \to \Ltwo \mathcal{U}(\Alg{S}/ R)$ such that the outer square in the diagram below on the left commutes.

$$
\xymatrix@C=0.3cm@R=-0.pc{
{\mathcal{U}\Alg{S}} \ar[dddd]_{\mathcal{U}c} \ar@{->>}[rrrr]^{\mathcal{U}[-]_R} &&& & {\mathcal{U}(\Alg{S}/ R)} \ar@{.>}[dddd]_{c_R} \ar@/^7ex/[dddddddd]^(0.5)f \\
\\
\\
\\
{\mathcal{U}\High \Alg{S} \raisebox{-7pt}{\rule{0pt}{20pt}}}  \ar@{>->}[dddd]_{e_{\Alg{S}}} \ar[rrrr]^{\mathcal{U}\High[-]_R} &&&&  {\mathcal{U}\High(\Alg{S}/ R) \raisebox{-7pt}{\rule{0pt}{20pt}}} \ar@{>->}[dddd]_{e_{\Alg{S}/R}}\\
\\
\\
\\
{\Ltwo \mathcal{U}\Alg{S}}  \ar[rrrr]_{\Ltwo\mathcal{U}[-]_R} &&&&  {\Ltwo \mathcal{U}(\Alg{S}/ R)} 
}
\quad
\xymatrix@C=0.3cm@R=0.1cm{
\mathcal{U}\Alg{S} \ar[dddddddd]_{\mathcal{U}\High[-]_R \circ \mathcal{U}c} \ar@{->>}[rrrr]^{\mathcal{U}[-]_R} &&& & \mathcal{U}(\Alg{S}/ R) \ar[dddddddd]^f \ar@{.>}[ddddddddllll]_{c_R}\\
\\
\\
\\
\\
\\
\\
\\
{\mathcal{U}\High(\Alg{S}/ R)\,\,}\ar@{>->}[rrrr]_{e_{\Alg{S}/R}} &&&& \Ltwo \mathcal{U}(\Alg{S}/ R)
}
$$

The bottom square commutes by naturality of $e$. The function $c_R$ is obtained by the epi-mono factorisation structure on $\Sets$ as shown on the right. To conclude that $R$ is a kernel bisimulation on $(\Alg{S},c)$, one only needs to show that $c_R$ is a map in $\EM(\T)$.

Let $\alpha \colon \T\mathcal{U}\Alg{S} \to \mathcal{U}\Alg{S}$ denote the algebra structure of $\Alg{S}$, that is $\Alg{S}= (\mathcal{U}\Alg{S},\alpha)$. 
Similarly $\Alg{S}/R= (\mathcal{U}(\Alg{S}/R),\alpha_R)$, $\High \Alg{S} = (\mathcal{U} \High \Alg{S} , \alpha_H) $ and $\High (\Alg{S}/R) = (\mathcal{U} \High( \Alg{S} /R) , \alpha_{RH})$. 
$$
\mkern150mu\xymatrix@C=0.3cm@R=0.3cm{
\T\mathcal{U}\Alg{S} \ar[drr]^{\alpha} \ar[dddd]_{\T\mathcal{U}c} \ar@{->>}[rrrr]^{\T\mathcal{U}[-]_R} && &&\T \mathcal{U}(\Alg{S}/ R) \ar'[d][dddd]^{\T c_R} \ar[drr]^{\alpha_R}\\
 && \mathcal{U}\Alg{S} \ar[dddd]_{\mathcal{U}c} \ar@{->>}[rrrr]^(0.3){\mathcal{U}[-]_R} && && \mathcal{U}(\Alg{S}/ R) \ar[dddd]^{c_R} \\
&&&&&&&&&&&&&&&&&&&&&&&&&&&&&&&&&&&&\\
&&&&&&&&&&&&&&&&&&&&&&&&&&&&&&&&&&&&\\
\T\mathcal{U}\High \Alg{S} \ar[drr]_{\alpha_H} \ar'[rr][rrrr]_{\T\mathcal{U}\High[-]_R} && &&  \T\mathcal{U}\High(\Alg{S}/ R) \ar[drr]^{\alpha_{RH}} \\
&& \mathcal{U}\High \Alg{S}  \ar[rrrr]_{\mathcal{U}\High[-]_R} && &&  \mathcal{U}\High(\Alg{S}/ R)
}
$$

Consider the above cube on $\Sets$.
The front face commutes by construction of $c_R$. The back face commutes as it is just $\T$ applied to the front face. The top face commutes because $R$ is a congruence: the map $[-]_R \colon (\mathcal{U}\Alg{S},\alpha) \to (\mathcal{U}(\Alg{S}/R),\alpha_R)$ is an $\T$-algebra homomorphism. Since $\High$ is a functor also  $\High [-]_R \colon \High(\mathcal{U}\Alg{S},\alpha) \to \High(\mathcal{U}(\Alg{S}/R),\alpha_R)$ is an $\T$-algebra homomorphism: this means that also the bottom face commutes. The leftmost face commutes since, by assumption, $c$ is a homomorphism of $\T$-algebras. 

To prove that also $c_R$ is a homomorphism of $\T$-algebras amounts to checking that also the rightmost face commutes. For this it is essential that $\T$ preserves epis, as every $\Sets$-endofunctor does, so that $\T\mathcal{U}[-]_R$ is an epi. From this fact and the following derivation, we conclude that $c_R \circ \alpha_R = \alpha_{RH} \circ \T c_R $.
\[
  \begin{array}[b]{rcl}
c_R \circ \alpha_R \circ \T\mathcal{U}[-]_R & = & c_R \circ \mathcal{U}[-]_R \circ \alpha \\
& = &  \mathcal{U}\High[-]_R \circ \mathcal{U}c \circ \alpha \\
& = & \mathcal{U}\High[-]_R \circ  \alpha_H \circ \T\mathcal{U}c \\
& = & \alpha_{RH} \circ \T\mathcal{U}\High[-]_R   \circ \T\mathcal{U}c \\
& = & \alpha_{RH} \circ \T c_R   \circ \T\mathcal{U}[-]_R \hfill 
  \end{array}
  \tag*{\qedhere}
\]
\end{proof}

\noindent In particular,  
Proposition~\ref{prop:epi}, the fact that behavioural equivalence $\approx$ is a kernel bisimulation, and the following result ensure that the functor $\Ubar\colon \CoAlg_{\EM(\Dis)}{(\PowC+1)^L} \to \CoAlg_{\Sets}{\Pow^L}$ preserves and reflects $\approx$.

\begin{prop}\label{prop:congruence}
Let $(\Alg{S},c)$ be a $(\PowC+1)^L$-coalgebra.
Behavioural equivalence on $\Ubar(\Alg{S},c)$ is a convex congruence\footnote{Convex congruences are congruences of convex algebras, see e.g.~\cite{SW15}. They are convex equivalences, i.e., closed under componentwise-defined convex combinations.}. Hence, $\Ubar$ preserves and reflects behavioural equivalence.
\end{prop}

\begin{proof}
$\Ubar(\Alg{S},c)$ is a coalgebra for the functor $\Pow^L \colon \Sets \to \Sets$, namely a labeled transition system. It is well known that for this kind of coalgebras,  behavioural equivalence ($\approx$) coincides with the standard notion of bisimilarity~\cite{Rut00:tcs}. 

We can thus proceed by exploiting coinduction and prove that the following relation is a bisimulation (in the standard sense). 
$$R = 
\{ (\psum{p}{\zeta_1}{\xi_1},\psum{p}{\zeta_2}{\xi_2}) \mid \zeta_1\approx \zeta_2, \;\; \xi_1 \approx \xi_2 \text{ and } p\in[0,1] 
\}$$
Suppose that $\psum{p}{\zeta_1}{\xi_1} \Tr{a} \epsilon_1$. Then, by Proposition \ref{prop:key}, $\epsilon_1 = \psum{p}{\zeta_1'}{\xi_1'}$ with $\zeta_1\Tr{a} \zeta_1'$ and $\xi_1 \Tr{a}\xi_1'$. Since $\zeta_1\approx \zeta_2$ and $\xi_1 \approx \xi_2$, and since $\approx$ is (standard) bisimilarity, there exist $\zeta_2' \approx \zeta_1'$ and $\xi_2' \approx \xi_1'$ such that 
$\zeta_2 \Tr{a} \zeta_2'$ and $\xi_2 \Tr{a}\xi_2'$. Again, by Proposition  \ref{prop:key}, $\psum{p}{\zeta_2}{\xi_2} \Tr{a}  \psum{p}{\zeta_2'}{\xi_2'}$. Moreover, by definition of $R$, $\psum{p}{\zeta_1'}{\xi_1'} \;R \; \psum{p}{\zeta_2'}{\xi_2'}$.
\end{proof}

%
This means that $\approx$ for $(\PowC+1)^L$-coalgebras, called transition systems on convex algebras in Section \ref{ssec:TSCA}, coincides with the standard notion of bisimilarity for LTSs.

\medskip

Table~\ref{tab:PA-coalgebras} summarises all models of PA: from the classical model $M$ being a $(\Pow\Dis)^L$-coalgebra $(S, c_M)$ on $\Sets$, via the convex model $M_c$ obtained as $\mathcal T_{\con}(S, c_M)$, to the belief-state transformer $M_{bs}$ 
that coincides with $\Ubar\after \Fbar \after \mathcal T_{\con}(S, c_M)$. 
The first line recalls the concrete notation for each model, the second the corresponding $\Sets$-coalgebra type, the third the way it is related to the original model $M$, and the last spells out the definition of the transition function in relation to the transition function $c_M$ of $M$.

\begin{table}
	\begin{center}
    \def\arraystretch{1.2} 
	\begin{tabular}{@{}c@{\qquad}c@{\qquad}c@{}}
		\toprule
		$M = (S, \Act, \tr{})$ & $M_c = (S, \Act, \to_c)$ & $M_{bs} = (\Dis S, \Act, \Btr{})$\\
		$(S, c_M\colon S \to (\Pow\Dis)^L)$ & $(S, \overline{c}_M\colon S \to (C+1)^L)$ & $(\Dis S, \hat c_M\colon \Dis S \to (\Pow\Dis S)^L)$\\
		$(S, c_M)$ & $(S,\bar c_M) = \mathcal T_{\con} (S, c_M)$ & $(S, \hat c_M) = \Ubar \after \Fbar\after \mathcal T_{\con} (S, c_M)$\\
		$c_M$ & $\bar c_M = \con^L \after c_M$ & $\hat c_M = (e_{\mathcal{F}S} + 1)^L \circ \mathcal{U}\overline{c}_M^\sharp$\\\bottomrule 
	\end{tabular}
	\caption{The three PA models, their corresponding $\Sets$-coalgebras, and relations to $M$.\label{tab:PA-coalgebras}}
\end{center}
\end{table}
%
\begin{thm}\label{thm:distr-bis-is-beh-eq}
Let $(S,c_M)$ be a probabilistic automaton. For all $\xi,\zeta\in \Dis S$, \begin{center}
$\xi \sim_d \zeta \quad \Leftrightarrow \quad \xi \approx \zeta \,\,\text{~in~} \,\,\Ubar\after \GPC \after \mathcal T_{\con}(S,c_M)$. 
\end{center}
\end{thm}

\begin{proof}
We have that $\Ubar\after \GPC \after \mathcal T_{\con}(S,c_M) = (\Dis S, \hat c_M) = (\Dis S, (e_{\mathcal{F}S} + 1)^L \circ \mathcal{U}\overline{c}_M^\sharp)$.
Since behavioural equivalence in $(\Dis S, (e_{\mathcal{F}S} + 1)^L \circ \mathcal{U}\overline{c}_M^\sharp)$ is bisimilarity for LTSs, by Proposition~\ref{prop:congruence}, all that we need to show is that $(\Dis S, \hat c_M) = (\Dis S, (e_{\mathcal{F}S} + 1)^L \circ \mathcal{U}\overline{c}_M^\sharp)$ is exactly the belief-state transformer $M_{bs}$ induced by $M$, as announced by Table~\ref{tab:PA-coalgebras}. Moreover, since $(e_{\mathcal{F}S} + 1)^L$ is the identity embedding, it suffices to understand $\mathcal{U}\overline{c}_M^\sharp$.

First, by definition, $\mathcal T_{\con}(S,c_M) = (S, \bar c_M) = (S, \con^L \after c_M)$, and concretely,  $$s\tr{l}_c \xi \text{~~in~~} M_c \text{~~iff~~}  \xi \in \bar{c}_M(s)(l) = \con(c_M(s)(l)).$$

Then $\mathcal{U}\bar{c}_M^\sharp = \alpha \after \Dis \bar{c}_M \colon \Dis S \to \Dis((CS +1)^L) \to (CS +1)^L$ by Lemma~\ref{lem:GPC-quasi}, where $\alpha$ is the algebra structure from Example~\ref{ex:Pc-on-free}. Concretely, we have, for a distribution\footnote{In order to avoid confusion between convex operations and distributions, for the moment we use this explicit notation for distributions.} $[s_i \mapsto p_i]$ 
\begin{eqnarray*}
\mathcal{U}\bar{c}_M^\sharp ([s_i \mapsto p_i])(l)  & = & \alpha \after \Dis \bar{c}_M ([s_i \mapsto p_i])(l)\\
& = & \alpha (\Dis \bar{c}_M ([s_i \mapsto p_i])(l)\\
& = & \alpha ([\bar c_M(s_i) \mapsto p_i])(l) \\
& = & (\dagger)
\end{eqnarray*}

Now $[\bar c_M(s_i) \mapsto p_i]$ is the distribution that assigns probability $p_i$ to the function $l \mapsto \{\xi_i \mid s_i \tr{l}_c \xi_i\}$ if $s_i \tr{l}_c$, and  $l \mapsto *$ otherwise.
Hence, by the convex operations corresponding to the algebra structure  $\alpha$ from Example~\ref{ex:Pc-on-free} we have
\begin{eqnarray*}
	(\dagger) & = & \left\{\ensuremath{\begin{array}{ll}
	\sum p_i \{\xi_i \mid s_i \tr{l}_c \xi_i\}\,\,, & \forall i.\,s_i \tr{l}_c \\
	*\,\,, & \exists i.\,s_i \not\tr{l}_c \end{array}}\right.\\
	& = & \left\{\ensuremath{\begin{array}{ll}
	\{\sum p_i \xi_i \mid s_i \tr{l}_c \xi_i\}\,\,, & \forall i.\,s_i \tr{l}_c \\
	*\,\,, & \exists i.\,s_i \not\tr{l}_c
\end{array}}\right.
\end{eqnarray*}
which means\footnote{Coming back to the usual notation for distributions, as formal sums.} that $\sum p_i s_i = [s_i  \mapsto p_i] \Tr{l} \sum p_i \xi_i$ in  $(\Dis S, \hat c_M)$
whenever $s_i \tr{l}_c \xi_i$ which is exactly the same condition as $\sum p_i s_i \Tr{l} \sum p_i \xi_i$ in $M_{bs}$. This completes the proof that $M_{bs}$ is exactly the coalgebra $(\Dis S, \hat c_M)$.
\end{proof}

Hence, distribution bisimilarity is indeed behavioural equivalence on the belief-state transformer and it coincides with standard bisimilarity. 
\section{Bisimulations Up-To Convex Hull}\label{sec:bis-upto-cc}
As we mentioned in Section~\ref{sec:GPC}, the generalised determinisation allows for the use of up-to techniques~\cite{milner1989communication,PS11}. An important example is shown in~\cite{bonchi2013checking}: given a non-deterministic automaton $c\colon S \to 2\times \Pow(S)^L$, one can reason on its determinisation $\mathcal{U}c^\sharp \colon \Pow (S) \to 2\times \Pow(S)^L$ up-to the algebraic structure carried by the state space $\Pow (S)$. Given a probabilistic automaton $(S,\Act, \to )$, we would like to exploit the algebraic structure carried by $\Dis(S)$ to prove properties of the corresponding belief-states transformer $(\Dis(S), \Act, \Btr{})$. Unfortunately, the lack of a suitable distributive law~\cite{VW06} makes it impossible to reuse the abstract results in~\cite{bonchi2014coinduction}. Fortunately, we can redo all the proofs by adapting the theory in~\cite{PS11} to the case of probabilistic automata.

\medskip

Hereafter we fix a PA $M= (S,\Act, \to )$ and the corresponding belief-states transformer $M_{bs}=(\Dis(S), \Act, \Btr{})$. We denote by $Rel_{\Dis(S)}$ the lattice of
relations over $\Dis(S)$  and define the monotone function $\bis\colon Rel_{\Dis(S)} \to Rel_{\Dis(S)}$ mapping every relation $R \in Rel_{\Dis(S)}$ into
$$\begin{array}{lclcll}
\bis(R)&::= & \{(\zeta_1,\zeta_2) & \mid & \forall a\in L, & \forall \zeta_1' \text{ s.t. } \zeta_1 \Btr{a} \zeta_1',\, \exists  \zeta_2' \text{ s.t. } \zeta_2 \Btr{a} \zeta_2' \text{ and } (\zeta_1',\zeta_2')\in R, \\
 && &&& \forall \zeta_2' \text{ s.t. } \zeta_2 \Btr{a} \zeta_2',\, \exists  \zeta_1' \text{ s.t. } \zeta_1 \Btr{a} \zeta_1' \text{ and } (\zeta_1',\zeta_2')\in R  \}\text{.}
\end{array}$$
A \emph{bisimulation} is a relation $R$ such that $R\subseteq \bis(R)$. Observe that these are just regular bisimulations for labeled transition systems and that the greatest fixpoint of $\bis$ coincides exactly with $\BSTeq$.
The \emph{coinduction} principle informs us that to prove that $\zeta_1 \BSTeq \zeta_2$ it is enough to exhibit a bisimulation $R$ such that $(\zeta_1,\zeta_2)\in R$.

\begin{exa}\label{ex:BTSeq}
Consider the PA in Figure \ref{fig:weighted} (left) and the belief-state transformer generated by it (right). It is easy to see that the (Dirac distributions of the) states $x_2$ and $y_2$ are in $\BSTeq$:  the relation $\{(x_2,y_2)\}$ is a bisimulation. Also  $\{(x_3,y_3)\}$ is a bisimulation: both $x_3 \not \Btr{a}$ and $y_3 \not \Btr{a}$. More generally, for all $\zeta,\xi \in \Dis(S), \, p,q\in[0,1]$,  
$\psum{p}{\zeta}{x_3} \BSTeq \psum{q}{\xi}{y_3}$  since both 
\begin{equation}\label{eq:x3y3bad}
\psum{p}{\zeta}{x_3} \not \Btr{a} \text{ and } \psum{q}{\xi}{y_3} \not \Btr{a}\text{.}
\end{equation}
Proving that $x_0 \BSTeq y_0$ is more complicated. We will show this in Example \ref{exuptocvx} but, for the time being, observe that one would need an infinite bisimulation containing the following pairs of states.
%
  \begin{equation*}
    \xymatrix @C=1.3em @R=1.1em{ %
      x_0\ar@{|->}[r]^{a} \ar@{.}[d]& %
      x_1\ar@{|->}[r]^(0.4){a} \ar@{.}[d]& %
      \frac{1}{2}x_1+\frac{1}{2}x_2\ar@{|->}[r]^{a} \ar@{.}[d]& %
      \frac{1}{4}x_1+\frac{3}{4}x_2 \ar@{|->}[r]^(0.7){a} \ar@{.}[d] & %
      \dots \ar@{.}[d] \\ %
      y_0\ar@{|->}[r]^(0.35){a} & %
      \frac{1}{2}y_1+\frac{1}{2}y_2\ar@{|->}[r]^{a} & %
      \frac{1}{4}y_1+\frac{3}{4}y_2 \ar@{|->}[r]^(0.5){a} & %
      \frac{1}{8}y_1+\frac{7}{8}y_2 \ar@{|->}[r]^(0.7){a}& %
      \dots }
  \end{equation*}
Indeed, all the distributions depicted above have infinitely many possible choices for $\Btr{a}$ but, whenever one of them executes a depicted transition, the corresponding distribution is forced, because of \eqref{eq:x3y3bad}, to also choose the depicted transition.  
  \end{exa}
An \emph{up-to technique} is a monotone map $\f\colon Rel_{\Dis(S)} \to Rel_{\Dis(S)}$, while a \emph{bisimulation up-to} $\f$ is a relation $R$ such that $R\subseteq \bis \f(R)$. An up-to technique $\f$ is said to be \emph{sound} if, for all $R\in Rel_{\Dis(S)}$, $R\subseteq \bis \f(R)$ entails that $R\subseteq \,\BSTeq$. It is said to be \emph{compatible} if $\f \bis(R) \subseteq \bis \f(R)$. In \cite{PS11}, it is shown that every compatible up-to technique is also sound.

Hereafter we consider the \emph{convex hull} technique $\cvx \colon Rel_{\Dis(S)} \to Rel_{\Dis(S)}$ mapping every relation $R \in Rel_{\Dis(S)}$ into its convex hull which, for the sake of clarity, is 
\begin{equation*}
\cvx(R) =  \{(\psum{p}{\zeta_1}{\xi_1}, \psum{p}{\zeta_2}{\xi_2})  \mid  (\zeta_1,\zeta_2)\in R, \, (\xi_1,\xi_2) \in R \text{ and } p\in [0,1] \}\text{.}
\end{equation*}

\begin{prop}\label{prop:cvxcomp}
$\cvx$ is compatible. 
\end{prop}

\begin{proof}
Assume that $(\epsilon_1,\epsilon_2) \in \cvx (\bis(R))$. By definition of $\cvx$ there exist $p\in [0,1]$, $\zeta_1,\xi_1,\zeta_2,\xi_2\in \Dis(S)$ such that
\begin{enumerate}[label={\rm(\alph*)},itemsep=1mm]
\item $\epsilon_1 = \psum{p}{\zeta_1}{\xi_1}$, $\epsilon_2 =  \psum{p}{\zeta_2}{\xi_2}$, 
\item $(\zeta_1,\zeta_2)\in \bis(R)$  and  $(\xi_1,\xi_2) \in \bis(R)$. 
\end{enumerate}
To prove that $(\epsilon_1,\epsilon_2) \in \bis (\cvx(R))$, assume that $\epsilon_1 \Btr{a}\epsilon_1'$ for some $a\in A$ and $\epsilon_1'\in \Dis(S)$. Then, by (a) and Proposition~\ref{prop:key}, $\zeta_1\Btr{a}\zeta_1'$ and $\xi_1\Btr{a}\xi_1'$ and $\epsilon_1' =\psum{p}{\zeta_1'}{\xi_1'} $. By (b), there exist $\zeta_2',\xi_2' \in \Dis(S)$ such that 
\begin{enumerate}[label={\rm(\alph*)},itemsep=1mm]
\setcounter{enumi}{2}
\item $\zeta_2 \Btr{a} \zeta_2'$,  $\xi_2 \Btr{a} \xi_2'$ and 
\item $(\zeta_1',\zeta_2')\in R$,  $(\xi_1',\xi_2')\in R$. 
\end{enumerate}
From (c) and Proposition~\ref{prop:key}, it follows that $\epsilon_2 \Btr{a} \psum{p}{\zeta_2'}{\xi_2'}$. From (d), one obtains that $(\epsilon_1' = \psum{p}{\zeta_1'}{\xi_1'}, \psum{p}{\zeta_2'}{\xi_2'}) \in \cvx (R)$.

One can proceed symmetrically for $\epsilon_2 \Btr{a}\epsilon_2'$. 

Therefore, by definition of $\bis$, $(\epsilon_1,\epsilon_2)\in  \bis (\cvx(R))$.
\end{proof}

This result has two consequences: first, $\cvx$ is sound\footnote{In~\cite{SangiorgiV16} a similar up-to technique called ``up-to lifting" is defined in the context of probabilistic $\lambda$-calculus and proven sound.} and thus one can prove $\BSTeq$ by means of bisimulation up-to $\cvx$; second, $\cvx$ can be effectively combined with other compatible up-to techniques~\cite{PS11}. In particular, by combining $\cvx$ with up-to equivalence---which is well known to be compatible---one obtains up-to congruence $\cgr\colon Rel_{\Dis(S)} \to Rel_{\Dis(S)}$. This technique maps a relation $R$ into its congruence closure: the smallest relation containing $R$ which is a congruence.

\begin{prop}\label{prop:cgrcomp}
$\cgr$ is compatible. 
\end{prop}

The proof of Proposition~\ref{prop:cgrcomp} can be elegantly presented using the modular approach developed in~\cite{PS11} that we recall hereafter.

\medskip

Up-to techniques can be combined in a number of interesting ways. For a map $\f\colon \mathit{Rel}_{\Dis(S)}\rightarrow \mathit{Rel}_{\Dis(S)}$, the $n$-iteration of $\f$ is defined as $\f^{n+1}=\f\circ \f^n$ and $\f^0=\ide$, the identity function. The omega iteration is defined as $\f^\omega(R)=\bigcup_{i=0}^\infty \f^i(R)$. 
The following result from \cite{PS11} informs us that compatible up-to techniques can be composed resulting in other compatible techniques.
\begin{lem}
  \label{lemmacompositionality}
  The following functions are compatible:
  \begin{itemize}[itemsep=1mm]
  \item $\ide$: the identity function; 
  \item $\f\circ \g$: the composition of compatible
    functions $\f$ and $\g$;
  \item $\bigcup F$: the pointwise union of an arbitrary family $F$
    of compatible functions: $\bigcup F (R) = \bigcup_{\f\in
      F}\f(R)$;
  \item $\f^{\omega}$: the (omega) iteration of a compatible function
    $\f$, defined as $\f^\omega(R) = \bigcup_{i=0}^\infty \f^i(R)$
  \qed
  \end{itemize}
\end{lem}

\noindent Apart from $\cvx$, we are interested in the following up-to techniques. 
\begin{itemize}[itemsep=1mm]
\item the constant function $\rfl$ mapping every $R$ into the identity relation $Id\subseteq \Dis(S)\times \Dis(S)$;
\item the square function $\trans$ mapping every $R$ into $t(R) = R^2 = \{(\zeta_1,\zeta_3) \mid \exists \zeta_2.\,\, \zeta_1 R \zeta_2 R \zeta_3\}$;
\item the opposite function $\sym$ mapping every $R$ into its opposite relation $R^{-1}$.
\end{itemize}
It is easy to check that all these functions are compatible. Lemma~\ref{lemmacompositionality} allows us to combine them so to obtain novel compatible up-to techniques.
For instance the equivalence closure $\e \colon \mathit{Rel}_{\Dis(S)}\rightarrow \mathit{Rel}_{\Dis(S)}$ can be decomposed as $(\ide \cup \rfl \cup \trans \cup \sym)^\omega$. The fact that $\e$ is compatible follows immediately from Lemma~\ref{lemmacompositionality}. 

In a similar way, we get that $\cgr$ is compatible.

\begin{proof}[Proof of Proposition~\ref{prop:cgrcomp}]
Observe that $\cgr = (\ide \cup \rfl \cup \trans \cup \sym \cup \cvx)^\omega$. We know that $\rfl,\sym,\trans$ and $\cvx$ (Proposition \ref{prop:cvxcomp}) are compatible. Compatibility of $\cgr$ follows by Lemma \ref{lemmacompositionality}.
\end{proof}

  Since $\cgr$ is compatible and thus sound, we can use bisimulation up-to $\cgr$ to check $\BSTeq$.

  \begin{exa}\label{exuptocvx}
  We can now prove that, in the PA depicted in Figure \ref{fig:weighted}, $x_0 \BSTeq y_0$.
  It is easy to see that the relation $R=\{(x_2,y_2), (x_3,y_3), (x_1,\frac{1}{2}y_1+\frac{1}{2}y_2), (x_0,y_0)\}$ is a bisimulation up-to $\cgr$:
  consider $(x_1,\frac{1}{2}y_1+\frac{1}{2}y_2)$ (the other pairs are trivial) and observe that   
  $$\xymatrix @C=1.3em @R=1.1em{ %
      x_1\ar@{|->}[r]^(0.4){a} \ar@{.}[d]|R& %
      \frac{1}{2}x_1+\frac{1}{2}x_2\ar@{.}[d]|{{ \cgr}(R)} \\ %
     \frac{1}{2}y_1+\frac{1}{2}y_2\ar@{|->}[r]_{a} & %
      \frac{1}{4}y_1+\frac{3}{4}y_2 }
    \quad
    \xymatrix @C=1.3em @R=1.3em{ %
  x_1\ar@{|->}[r]^(0.4){a} \ar@{.}[d]|R& %
   \frac{1}{2}x_3+\frac{1}{2}x_2\ar@{.}[d]|{{ \cgr}(R)} \\ %
      \frac{1}{2}y_1+\frac{1}{2}y_2\ar@{|->}[r]_{a} & %
      \frac{1}{2}y_3+\frac{1}{2}y_2  }
  $$
  
Since all the transitions of $x_1$ and $\frac{1}{2}y_1+\frac{1}{2}y_2$ are obtained as convex combination of the two above, the arriving states are forced to be in $\cgr (R)$. In symbols, if $$x_1\Btr{a}  \zeta = \psum{p}{\left(\frac{1}{2}x_1+\frac{1}{2}x_2\right)}{\left(\frac{1}{2}x_3+\frac{1}{2}x_2\right)},$$ then  $$\frac{1}{2}y_1+\frac{1}{2}y_2 \Btr{a} \xi = \psum{p}{ \left( \frac{1}{4}y_1+\frac{3}{4}y_2\right)}{\left(\frac{1}{2}y_3+\frac{1}{2}y_2\right)}$$ and $(\zeta,\xi)\in \cgr (R) $.  
\end{exa}

Recall that in Example \ref{ex:BTSeq}, we showed that  to prove $x_0 \BSTeq y_0$ without up-to techniques one would need an infinite bisimulation. Instead, the relation $R$ in Example \ref{exuptocvx} is a \emph{finite} bisimulation up-to $\cgr$. It turns out that one can always check $\BSTeq$ by means of only finite bisimulations up-to. The key to this result is the following theorem, proved in~\cite{SW15}.

\begin{thm}
Congruences of finitely generated convex algebras are finitely generated. \qed
\end{thm} 

This result informs us that for a PA with a finite state space $S$, $\BSTeq\,\,\subseteq\,\, \Dis(S)\times \Dis(S)$ is finitely generated (since $\BSTeq$ is a congruence, see Proposition~\ref{prop:congruence}). In other words there exists a finite relation $R$ such that $\cgr(R)\,\,=\,\,\BSTeq$. Such $R$ is a finite bisimulation up-to $\cgr$:
\begin{equation*}
R\,\subseteq\, \cgr(R) \,=\, \BSTeq\, = \,\,\bis (\BSTeq)\, = \,\bis (\cgr(R))\text{.}
\end{equation*}

\begin{cor}\label{cor:finite}
Let $(S,\Act, \to )$ be a finite PA and let $\zeta_1,\zeta_2\in \Dis(S)$ be two distributions such that $\zeta_1 \BSTeq \zeta_2$. Then, there exists a finite bisimulation up-to $\cgr$ $R$ such that $(\zeta_1,\zeta_2)\in R$. \qed
\end{cor}

\noindent This last observation has another pleasant consequence\footnote{We thank Matteo Mio for pointing out this consequence to us.}: distribution bisimilarity $\BSTeq$ is decidable, provided we restrict to finite PA with rational probabilities. Indeed, given a finite PA $(S,\Act, \to )$ with rational probabilities, by Corollary~\ref{cor:finite}, checking distribution bisimilarity is semi-decidable (it requires enumerating all finite bisimulations up-to). At the same time, proving that two distributions are not distribution bisimilar is also enumerable, having a finite proof. Together, these two facts imply decidability of distribution bisimilarity.


\section{Conclusions and Future Work}\label{sec:conc}

Belief-state transformers and distribution bisimilarity have a strong coalgebraic foundation which leads to a new proof method---bisimulation up-to convex hull. More interestingly, and somewhat surprisingly, proving distribution bisimilarity can be achieved using only {\em finite} bisimulation up-to witness. This opens exciting new avenues: Corollary~\ref{cor:finite} gives us hope that bisimulations up-to may play an important role in designing algorithms for automatic equivalence checking of PA, similar to the one played for NDA. Exploring their connections with the algorithms in~\cite{HermannsKK14,EisentrautHKT013} is our next step.

From a more abstract perspective, our work highlights some limitations of the bialgebraic approach~\cite{TuriP97,Bartels04:thesis,Klin11}. Despite the fact that our structures are coalgebras on algebras, they are not bialgebras: but still $\approx$ is a congruence and it is amenable to up-to techniques. We believe that \emph{lax} bialgebra may provide some deeper insights.

Last but not least, yet fully orthogonal, is the question whether the determinisation-based semantics, in particular our distribution bisimilarity, allows for an abstract generalisation towards weak semantics. Weak semantics are very important in applications and are favoured in process theory. However, in coalgebra, weak semantics are difficult to treat as they lack an elegant abstract representation. There are several approaches to coalgebraic weak semantics without a clear consensus beyond the fact that weak semantics and coalgebra are not a perfect match. Having the monad structure on the set of states might be beneficial.


\bibliographystyle{alpha}
\bibliography{CAPower}

\end{document}